\documentclass{aa}

\usepackage{graphicx}
\usepackage[varg]{txfonts}
\usepackage[urlcolor=cyan, colorlinks=true, citecolor=blue, linkcolor=blue]{hyperref}
\usepackage{natbib}
\bibpunct{(}{)}{;}{a}{}{,} 
\usepackage{cleveref}
\usepackage[super]{nth}
\usepackage{upgreek}
\usepackage{ulem}
\usepackage{multirow}

\usepackage{tablefootnote}
\begin{document} 

\title{The first catalogue of spectroscopically confirmed red nuggets\\at z$\sim$0.7 from the VIPERS survey}
\subtitle{Linking high-z red nuggets and local relics} 

\author{Krzysztof Lisiecki\inst{1,2}, Katarzyna Małek\inst{2,3}, Małgorzata Siudek\inst{2,4,5}, Agnieszka Pollo\inst{2,6}, Janusz Krywult \inst{7}, Agata Karska\inst{8,1}, Junais\inst{2}}

\institute{
$^{1}$ Institute of Astronomy, Faculty of Physics,
Astronomy and Informatics, Nicolaus Copernicus University, ul. Grudziądzka 5, 87-100 Toruń, Poland 	\email{klisiecki@astro.umk.pl }\\
$^{2}$ National Centre for Nuclear Research, Pasteura 7, 093, Warsaw, Poland \\
$^{3}$ Aix Marseille Univ. CNRS, CNES, LAM, Marseille, France \\
$^{4}$ Institut de Física d'Altes Energies (IFAE), The Barcelona Institute of Science and Technology, E-08193 Bellaterra (Barcelona), Spain\\
$^{5}$ Institute of Space Sciences (ICE, CSIC), Campus UAB, Carrer de Magrans, 08193 Barcelona, Spain\\
$^{6}$ Astronomical Observatory of the Jagiellonian University, ul. Orla 171, 30-244 Kraków, Poland \\
$^{7}$ Institute of Physics, Jan Kochanowski University, ul. Swietokrzyska 15, 25-406, Kielce, Poland\\
$^{8}$ Max-Planck-Institut für Radioastronomie, Auf dem Hügel 69, 53121, Bonn, Germany\\
}

\date{Received 23 March 2022; Accepted 8 August 2022}
\titlerunning{Red nuggets at z$\sim$0.7 from VIPERS}
\authorrunning{K. Lisiecki et al., 2022}

  \abstract
{ 
'Red nuggets' are a rare population of passive compact massive galaxies thought to be the first massive galaxies that formed in the Universe. First found at $z \sim 3$, they are even less abundant at lower redshifts, and it is believed that with time they mostly transformed through mergers into today's giant ellipticals. Those red nuggets which managed to escape this fate can serve as unique laboratories to study the early evolution of massive galaxies. 
}
{
In this paper, we aim to make use of the unprecedented statistical power of the VIMOS Public Extragalactic Redshift Survey to build the largest up-to-date catalogue of spectroscopically confirmed red nuggets at the intermediate redshift $0.5<z<1.0$.}
{  
Starting from a catalogue of nearly 90\,000 VIPERS galaxies we select sources with stellar masses $M_{star} > 8\times10^{10}$ $\rm{M}_{\odot}$ and effective radii $R_\mathrm{e}<1.5$ kpc. 
Among them, we select red, passive galaxies with old stellar population based on colour--colour NUVrK diagram, star formation rate values, and verification of their optical spectra. 
}
{
Verifying the influence of the limit of the source compactness on the selection, we found that the sample size can vary even up to two orders of magnitude, depending on the chosen criterion.
Using one of the most restrictive criteria with additional checks on their spectra and passiveness, we spectroscopically identified only 77 previously unknown red nuggets.
The resultant catalogue of 77 red nuggets is the largest such catalogue built based on the uniform set of selection criteria above the local Universe.
Number density calculated on the final sample of 77 VIPERS passive red nuggets per comoving Mpc$^3$ increases from 4.7$\times10^{-6}$ at $z \sim 0.61$ to $9.8 \times 10^{-6}$ at $z \sim 0.95$,
which is higher than values estimated in the local Universe, and lower than the ones found at $z>2$.
It fills the gap at intermediate redshift.

}
{A catalogue of red nuggets presented in this paper is a golden sample for future studies of this rare population of objects at intermediate redshift.
Besides covering a unique redshift range and careful selection of galaxies, the catalogue is spectroscopically identified. 
}

\maketitle

\section{Introduction}
Morphological classification of galaxies splits them into two main groups, early and late types \citep{hubble26}.
The early-type galaxies (ETGs) are spheroidal objects with smooth brightness distributions, low star formation (SF) rates, and stellar content dominated by evolved stars.
The late-type galaxies are at the opposite side of Hubble's tuning fork diagram: those are mostly spiral galaxies and contain both young and evolved stars \citep{kennicut98}.

It has been shown that the most massive ETGs in the local Universe differ significantly from their counterparts at high redshifts ($z>2$).
These high-redshift ETGs are extremely compact, a few times smaller in size than their local counterparts \cite[][]{daddi2005,trujillo07,damjanov09,vanderwel14}.
According to the so-called two-phase formation scenario, during the initial phase dominated by numerous wet mergers between gas-rich galaxies and characterized by high star formation rate (SFR), progenitors of ETGs increase their stellar mass rapidly, even up to $M_{star} \sim 10^{11}$ M$_{\odot}$, but remain compact in size \citep{oser10}.
Subsequently, star formation in these compact and massive objects quenches quickly, and the galaxies become passive, compact, and massive. 
At this stage, those are referred to as a \lq red nuggets'.
In the second, final phase, red nuggets undergo dry mergers with other
gas-poor galaxies which results in the increase of their size and their transformation in giant elliptical galaxies \citep{vandokkum10}.
This two-phase scenario has been confirmed observationally \citep{barro13, zibetti20}, and is predicted by theoretical models and numerical simulations \citep{naab09, zolotov15, flores-freitas21}.

Due to the stochastic nature of mergers, some red nuggets skip the second phase of the two-phase scenario and end as so-called \lq relics' \citep{ferre-mateu17}.
Red nuggets (relics) provide a unique opportunity to study stellar populations that remained relatively unaltered for billions of years.
Yet, direct observational studies of red nuggets have been cursory so far due to their high-redshift and the lack of sufficient angular resolution.
Moreover, the probability of finding the relics in the local Universe is very low, as most of them already merged.
The high-resolution cosmological simulations predict that the fraction of $z \sim2$ red nuggets that evolve into local Universe relics is less than 15\% \citep{quilis13, wellons16, furlong17}.
These predictions vary depending on the exact description of physical processes influencing the galaxy evolution, in particular stellar winds or Active Galaxy Nuclei (AGN) feedback, in the simulations.

Despite numerous observational studies dedicated to identifying relics at the lowest redshifts, $z\leq 0.2$, their estimated number densities differ by a few orders of magnitude between different samples \citep{trujillo09, taylor10, valentinuzzi10, poggianti13,saulder15, tortora16,ferre-mateu17,scognamiglio20}. 
A likely reason for these differences is different selection criteria for compact sources applied by the authors.
At redshifts from $0.2$ to $0.5$, there are studies that showed similar number densities of relics or ultracompact massive galaxies (hereafter UCMGs), reporting many of these kinds of objects in systematic wide-field surveys \citep{tortora16, charbonnier17, buitrago18, scognamiglio20}.
However, the quantitive comparison of the number densities is not straightforward due to different selection compactness criteria.
Finally, at redshifts from $0.5$ to $3$, limited angular resolution and a lack of systematic wide-field spectroscopic surveys hinder the detection of passive UCMGs \citep{barro13, vanderwel14, vandokkum15}.

In the presented paper, we explore the galaxy sample from the VIMOS Public Extragalactic Redshift Survey \citep[VIPERS,][]{scodeggio18}, which is a spectroscopic survey of about~90\,000 galaxies at redshift $0.4 < z < 1.2$, to spectroscopically identify new red nuggets at intermediate redshift.
The availability of spectra provides a major improvement over most previous observational studies of red nuggets since it allows to confirm and precisely measure their redshifts, critical for estimating physical quantities of galaxies.
Moreover, our intermediate redshift sample offers a unique opportunity to bridge the gap between high-redshift red nuggets and their local counterparts.

The paper is organized as follows.
In Section~\ref{sec:data}, we present the VIPERS survey and observational data which are relevant for this work, in particular effective radii and stellar masses.
We also describe the initial sample selection criteria used for further analysis.
This section is followed by Sec.~\ref{sec:cigale}, with a discussion about our spectral energy distribution fitting procedure and reestimating stellar masses. 
The final selection of red nuggets, including compactness discrepancy and passiveness criteria is presented in Sec.~\ref{Sec:red_nuggets_selection}.
In Sec.~\ref{sec:properties} the properties of red nuggets are shown.
Finally, discussion and comparison with other results are provided in Section~\ref{sec:discussion}, which is followed by a summary in Section~\ref{Sec:summary}.

Throughout the paper, we assume the $\Lambda$CDM cosmological model with $H_0 = 70\rm{km s}^{-1} \rm{Mpc}^{-1}$, $\Omega_m = 0.3$ and $\Omega_\Lambda = 0.7$.

\section{Data}\label{sec:data}
The VIMOS Public Extragalactic Redshift Survey (hereafter VIPERS) is a completed ESO Large Program, which was designed to investigate the spatial distribution of galaxies over the $z$\textasciitilde1 Universe \citep{scodeggio18}. 
It extends over an area of 23.5~deg$^2$ and has provided a catalogue of spectroscopic redshifts for nearly 90\,000 galaxies.  
Spectroscopic targets were selected within the W1 and W4 fields of the Canada-France-Hawaii Telescope Legacy Survey Wide (CFHTLS-Wide) to a limit of $i$ < 22.5 mag, with a simple and robust ($r-i$) vs ($u-g$) colour--colour pre-selection to effectively remove galaxies at $z$ < 0.5. 
The spectra were observed using the VIMOS spectrograph \citep{lefevre03} with the LR Red grism, providing a wavelength coverage of 5\,500-9\,500~{\AA} with a resolution R $\simeq$ 220.
Taking into account volume and sampling,  VIPERS may be considered as the intermediate redshift ($z\sim0.7$) equivalent of state-of-the-art local surveys ($z$<0.2), such as the 2dF Galaxy Redshift Survey \cite[2dFGRS;][]{colless01} and Sloan Digital Sky Survey \cite[SDSS;][]{york00,ahmuda20}.

The quality of the VIPERS redshift measurement is quantified at the time of validation by attributing a redshift flag ($z_{flag}$).  
The $z_{flag}$  ranges from a value of 4, indicating >99\% of confidence that the redshift measurement is secure, to 0, corresponding to no redshift estimate.
In the following analysis, we consider only objects whose redshift measurement quality was larger than 95\%: $z_{flag} \in \{3, 4, 23, 24\}$ \citep{garilli14,scodeggio18}, where 23 and 24 stand for $>$95\% confidence of the measured spectroscopic redshift for serendipitous targets. 
A detailed description of the survey is given by \citet{guzzo14} and \citet{scodeggio18} and the specification of the pipeline used for data reduction with the quality flag system are described by \citet{garilli14}.

In the following analysis, in particular in number density estimations, we used also spectroscopic success rate (SSR) and target sampling rate (TSR).
Both parameters provide information about the completeness of the VIPERS parent catalogue  \cite[more details can be found in Sec.~1 of][]{scodeggio18}.
The SSR is the fraction of detected galaxy targets with a reliable spectroscopic redshift measurement and all the spectroscopic targets.
Only $\sim$45\% of available targets were assigned a slit.
For this reason, TSR is defined as the fraction of candidate galaxies for which spectrum has been acquired.
For detailed description, we refer to \cite{garilli14}.

The VIPERS spectroscopic data are accompanied by a wealth of ancillary information. 
In particular, in this work, we make use of the multi-wavelength photometric catalogue (see Sec.~\ref{sec:photometry}) and physical parameters derived via spectral energy distribution (hereafter SED) fitting by \citet{moutard16_II} (see Sec.~\ref{sec:stellar_masses}), as well as morphological parameters derived by \cite{krywult17} (see Sec.~\ref{sec:radius}).

\subsection{Photometric data}\label{sec:photometry}
Photometric data for this analysis has been taken from the VIPERS database \citep{moutard16_I} providing multi-wavelength observations from the ultraviolet (hereafter UV) to the infrared (IR) wavelengths. 
The catalogue combines CFHTLS T0007-based photometric measurements in \textit{u}, \textit{g}, \textit{r}, \textit{i}, \textit{y} (the filter \textit{i} broke in 2006 and it was replaced by a similar, but not identical filter, called \textit{iy}), and \textit{z} filters with GALEX FUV and NUV, and the CFHT WIRCam K$_s$-band observations, complemented by VISTA K photometry from the VIDEO survey \citep{jarvis13}.
In addition to the NIR and UV photometry, the catalogue provides the MIR photometry in the W1 field with Spitzer/IRAC channels (3.6, 4.5, 5.8 and 8.0 $\mu m$) and MIPS filters (24, 70, and 160~$\mu$m) from the Spitzer WIDE-area Infrared Extragalactic Survey (SWIRE). 
The VIPERS fields were also covered by NASA’s Wide-field Infrared Survey Explorer \cite[WISE;][]{wright10} passbands W1, W2, W3, and W4 with effective wavelengths 3.4, 4.6, 12.1, and 22.5~$\mu$m, respectively. 
Table~\ref{tab:photometric_data}  shows the photometric bands used for our data with its centered wavelength.

\begin{table}[ht] 
\centering
\caption{Summary of available photometric data in each band used for the SED fitting. The third column, $\lambda_{mean}$, is the centre of the specific filter band given in $\mu$m. The last column, N$_{gal}$, provides the number of detections among the sample of  6\,961 \textit{UCMG candidates}. }
    \begin{tabular}{l l c c}
       Telescope/ & \multirow{2}{*}{Filter} & $\lambda_{mean}$ & \multirow{2}{*}{N$_{gal}$} \\
      Instrument &          & \small{($\mu$m)}     & \\ \hline
GALEX       &  FUV          &\phantom{22}0.155 &  \phantom {2}\,348\\
            &  NUV          &\phantom{22}0.234 &  2\,352 \\
CFHT/MegaCam&  \textit{u}            &\phantom{22}0.369 &  6\,793 \\
            &  \textit{g}            &\phantom{22}0.482 &  6\,960 \\
            &  \textit{r}            &\phantom{22}0.643 &  6\,955 \\
            &  \textit{i}            &\phantom{22}0.772 &  5\,689 \\
            &  \textit{z}            &\phantom{22}0.900 &  6\,958 \\
            &  \textit{iy}           &\phantom{22}0.769 &  1\,272 \\
CFHT/Wircam &  K$_s$        &\phantom{22}2.150 &  \phantom {2}\,476 \\
VISTA       &  K$_{video}$  &\phantom{22}2.158 &  6\,445 \\
WISE        &  W1           &\phantom{22}3.353 &  4\,499 \\
            &  W2           &\phantom{22}4.603 &  4\,499 \\
            &  W3           &\phantom{2}11.561 &  4\,499 \\
            &  W4           &\phantom{2}22.088 &  4\,499 \\
Spitzer/IRAC&  I1           &\phantom{22}3.557 &  1\,299 \\
            &  I2           &\phantom{22}4.505 &  1\,334 \\
            &  I3           &\phantom{22}5.739 &  \phantom {2}\,411 \\
            &  I4           &\phantom{22}7.927 &  \phantom {2}\,118 \\
Spitzer/MIPS&  24$\mu$m     &\phantom{2}23.843 &  \phantom {2}\,219 \\
            &  70$\mu$m     &\phantom{2}72.555 &  \phantom{2\,2}17 \\
            &  160$\mu$m    &157.000 & \phantom {2\,22}4 \\
            
\hline
\end{tabular}
\label{tab:photometric_data}
\end{table}

\subsection{Stellar mass}\label{sec:stellar_masses}
Stellar masses, $M_{star}$, for VIPERS galaxies used in the initial sample selection have been derived by \citet{moutard16_II} using the stellar population synthesis models of \cite{bruzal03} with Le Phare \citep{arnouts99, ilbert06}.
Through the procedure of modelling, authors followed \cite{ilbert13}.
Values of stellar masses correspond to the median of the stellar mass probability distribution marginalised over all other fitted parameters.
It should be noted, that those stellar masses do not have uncertainties, which have an important impact on our further analysis.
For details see \cite{moutard16_II}.

\subsection{Effective radii and morphological parameters}\label{sec:radius}

The radial surface brightness profiles of the galaxies can be modelled by a S\'ersic profile \citep{sersic63, sersic68}:
\begin{equation}
    I(r) = I_e\exp{ \left( -b_n \left[ \left( \frac{r}{R_e} \right)^{1/n} -1 \right] \right)},
\end{equation}
where $R_e$ is the radius enclosing half of the total light of the galaxy, 
$I_e$ is the mean surface brightness at $R_e$ and $n$ is the S\'ersic index. The coefficient $b_n$ depends on the value of $n$ and is defined in such a way that $R_e$ encloses half of the total light \citep{graham2005}.

A single component 2D S\'ersic profile fitting was done by \citet{krywult17} on the CFHTLS {\it i}-band images of the VIPERS galaxies using GALFIT \citep{peng02}. Since GALFIT performs an elliptical isophotal fitting, it provides the semi-major axis ($a_e$) corresponding to the $R_e$ and an axis ratio ($b/a$) for a galaxy. Therefore, in the following analysis, we use the circularised half-light radius ($R_e = a_e\sqrt{b/a}$), to compare our results with other studies.

The goodness of the fit is measured by reduced $\chi^2$ and we refer to it as $\chi^2_{GALFIT}$.
The accuracy of structural parameters was derived from simulated galaxies on CFHTLS images, returning the uncertainties in effective radii measurements at the level of 4.4$\%$ for 68$\%$ of VIPERS sample, and at the level of 12$\%$ for 95$\%$ of VIPERS sample.
A~detailed description of VIPERS morphological parameters can be found in \cite{krywult17}.

\subsection{Initial sample selection}\label{sec:sample_cigale}

The VIPERS survey contains spectroscopic observations of 91\,507 sources, including stars and active galactic nuclei.
Aiming at building a VIPERS red nuggets sample we started from the selection of an initial \textit{pure sample}. 

\subsubsection{Secure redshift measurements and redshift range}
Firstly, we selected 54\,252 objects with secure redshift measurements with a confidence level higher than 95\% \citep[with $z_{flag} \in$ \{3, 4, 23, 24\}, see][for details]{garilli14}.
In the next step, we selected our sample based on the redshift range.
Narrowing the redshift range to $0.5 \leq z \leq 1$ due to the colour completeness of VIPERS survey \citep{fritz14} further reduced the sample to 44\,145 galaxies.

\subsubsection{Effective radii quality}
Following, we restricted this sample to 36\,635 sources, for which $R_e$ uncertainties were derived.
The absence of uncertainties in $R_e$ indicates a problem in the convergence of the fitting procedure \citep[for details see ][]{krywult17} and such objects are not suitable for this analysis.
We furthermore only considered sources with values of $\chi_{GALFIT}^2$ smaller than 1.2. Moreover, following \citet{krywult17}, we also removed the galaxies with a S\'ersic index $n<0.2$, as the estimated morphological properties for such sources may be unreliable.
The last criterion to select our \textit{pure sample} is the reliability of $R_e$ measurements with relative errors $<$100\%.
Finally, it gave us 36\,157 sources (hereafter \textit{pure sample}).

\subsubsection{Stellar mass and compactness}
In the next step, we performed further selection from the \textit{pure sample} by taking into account physical properties of observed objects, namely stellar masses, $M_{star}$ and effective radii, $R_e$, which are already available (see Sec.\ref{sec:stellar_masses} and Sec.\ref{sec:radius}).
The most important criteria to find the UCMGs is compactness, but the ambiguity in the literature is strong (see Tab.\ref{tab:compactness_cuts}).
All of these criteria are based on the position of selected objects in the $R_e$~vs~$M_{star}$ diagram.
The chosen threshold crucially affects the number of galaxies in the selected sample.
In order to maximize our sample of potential UCMGs, we first applied the most liberal threshold proposed by \citealt[][which original equation is shown in Table \ref{tab:compactness_cuts}]{damjanov15}.
As we do not have information about uncertainties of given stellar masses, we slightly modified the \citet{damjanov15} criterion by adding 0.1 dex to~$R_e$:\begin{equation}\label{damj-dex}
     (\log(R_e) + 5.74 -0.1)/\log(M_{star}/\rm{M}_\odot) < 0.568. 
\end{equation}
This was necessary to ensure that we do not remove any probable UCMGs from further analysis 
due to inaccurate stellar-mass estimates from Le Phare \citep{arnouts99, ilbert06}. This additional arbitrary change of 0.1 dex results in 2\,571 galaxies added to the final analysis. It corresponds to the $\sim$58\% increase in the number of analyzed galaxies compared to the original \cite{damjanov14} criterion.
While our 0.1 dex is arbitrary, as described in the following sections we found no red nuggets candidates in the $M_{star}$ -- $R_e$ space between the original \cite{damjanov14} criterion, and the one enlarged by 0.1 dex. 
This serves as an additional sanity check, that no UCMGs are hiding above \cite{damjanov14} selection.
This criterion narrows down the initial sample to  6\,961 galaxies (hereafter \textit{UCMG candidates}). 
The summary of all performed cuts until this point, their order and impact on the sample are presented in Table~\ref{tab:pure_sample_cuts}.
 
\subsubsection{Pure initial sample} 
\begin{figure*}[ht]
\centering
\includegraphics[width = 0.98\textwidth]{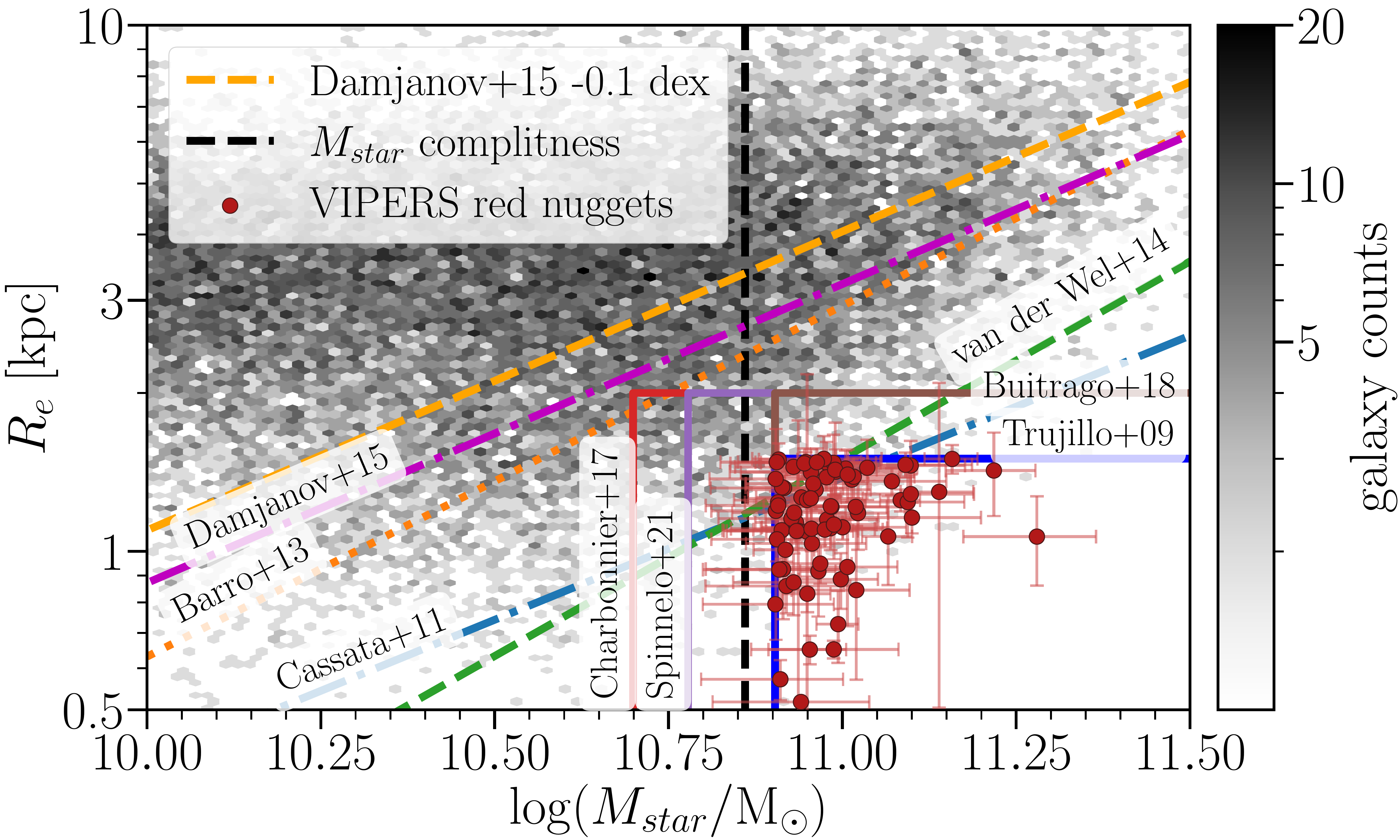}
\caption{Stellar mass vs effective radius distribution of 36\,157 galaxies within the \textit{pure sample}. The magenta dash-dotted line represents the initial \cite{damjanov15} cut for massive compact sources, the blue solid line indicates one of the most restrictive criteria proposed by \cite{trujillo09}, and the orange dashed line visualises the cut adopted in this work to select 6\,961 \textit{UCMG candidates}. The red points with error bars represent our final sample of 77 VIPERS red nuggets. The black dashed line shows the stellar mass completeness in the VIPERS catalogue.
With other lines different compactness criteria are marked:
blue dot-dashed line -- ultracompact \cite{cassata11};
orange dotted line -- \cite{barro13};
green dashed line -- ultracompact \cite{vanderwel14};
red solid line -- \cite{charbonnier17};
brown solid line -- \cite{buitrago18};
violet solid line -- \cite{spiniello21}.}
\label{fig:compactness}
\end{figure*}

The distribution of the VIPERS \textit{pure sample} in the $M_{star}$--$R_e$ plane is shown in Fig.~\ref{fig:compactness}. The orange dashed line represents the cut defined by Eq.~\ref{damj-dex}.
The original limit from \cite{damjanov15} is plotted with a magenta dash-dotted line.
The blue solid line shows one of the most conservative criteria for compactness found in the literature \citep{trujillo09} and red points show our final catalogue of 77 VIPERS red nuggets (details in Sec. \ref{Sec:red_nuggets_selection}).
 
\begin{figure}[ht]
\centering
\includegraphics[width = 0.49\textwidth]{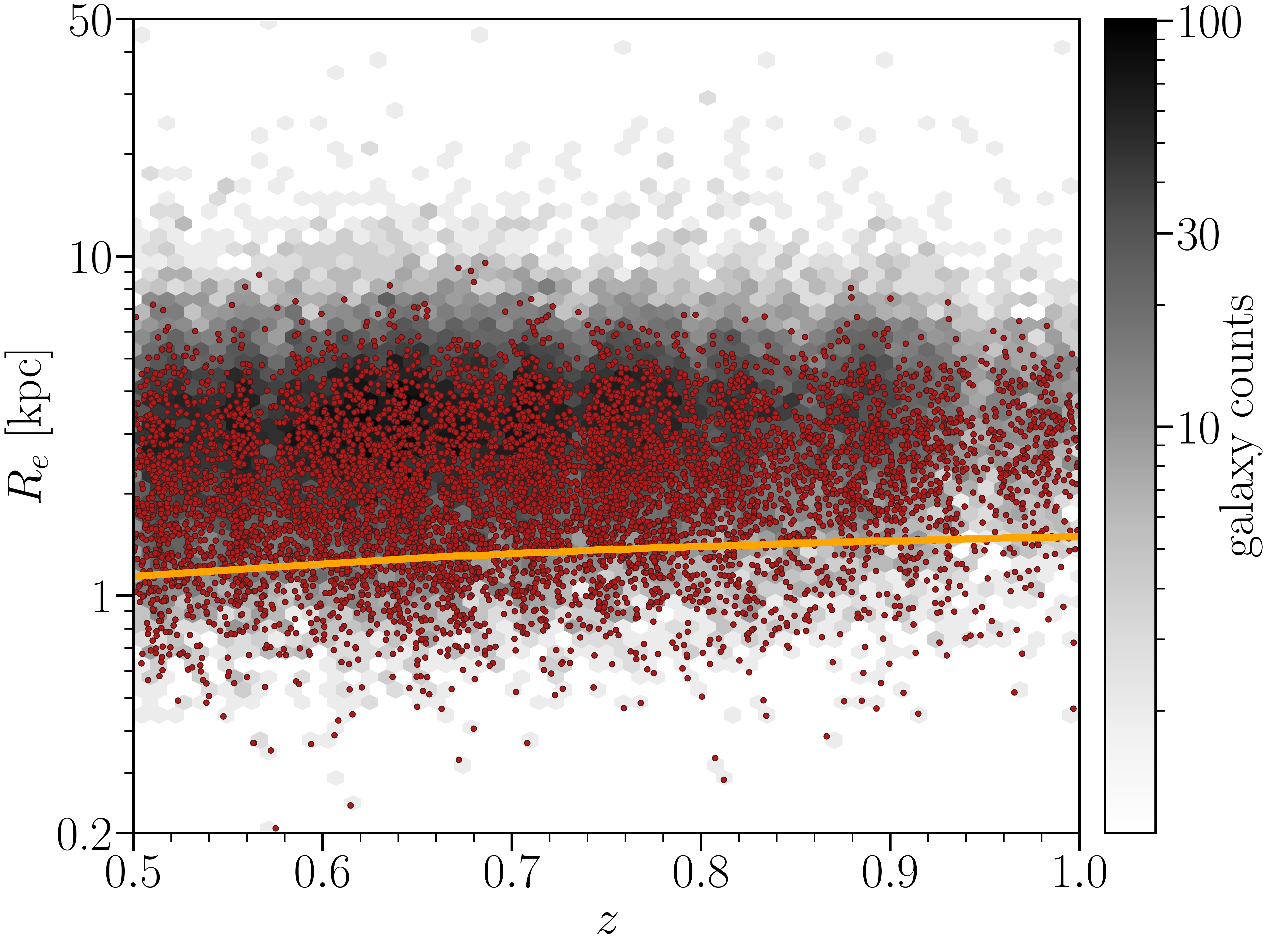}
\caption{Distribution of effective radius as a function of redshift for 36\,157 galaxies from VIPERS \textit{pure sample} is shown in grey. The orange line represents the VIPERS' seeing-detection limit. The \textit{UCMG candidates} are marked with red circles. }
\label{fig:Rz_pllt_with subsample}
\end{figure}

The orange solid line in Figure~\ref{fig:Rz_pllt_with subsample} represents the seeing-detection limit.
The CFHTLS $i$ images used by \cite{krywult17} have a pixel scale of 0.186$''$.
The mean seeing, defined by the Full Width at Half Maximum (FWHM) of stellar sources, depends on the CFHTLS filter.
According to \cite{goranowa09} it is equal 0.64$''$ for \textit{i}/\textit{iy}-band.
The angular size limit is transformed into physical units of $R_e$ as a function of redshift, which is shown as an orange solid line in Fig.~\ref{fig:Rz_pllt_with subsample}. 
A vast majority of both \textit{pure sample} (33\,810, $\sim$93.5\%), and \textit{UCMG candidates} (5\,673, $\sim$81\%) have sizes larger than the pixel size. 
We decided not to remove sources lying below the seeing-detection limit by default, but in the next steps of our analysis, we verified their properties more carefully (see Sec. \ref{Sec:red_nuggets_selection}).

\begin{table}[ht] 
\centering
\caption{Summary of initial cuts performed to select firstly \textit{pure sample}, and then \textit{UCMG candidates}.}
    \begin{tabular}{c c}
    \hline
    Cut & Sample size\\
    \hline
    VIPERS database                 & 91\,507 \\
    $z_{flag} \in \{3,4,23,24\}$      & 54\,252 \\
    Redshift range $0.5\leq z \leq 1$              & 44\,145 \\
    $R_e$ uncertainties and relative errors $<$ 100\%   & 36\,157 \\
    \textit{UCMG candidates} (based on Equation \ref{damj-dex})         & \phantom06\,961\\
\hline
\end{tabular}
\label{tab:pure_sample_cuts}
\end{table}

\section{Stellar mass re-estimation}\label{sec:cigale}

We decided to recalculate stellar masses of \textit{UCMG candidates} as a part of the quality check.
As the stellar mass is one of the two parameters which characterises compact sources, it is obligatory to take into account the goodness of the fit and uncertainties, which were not derived so far for VIPERS galaxies.
For this purpose, we used the state-of-art SED fitting tool called  Code Investigating GALaxy Emission \citep[CIGALE;][]{Boquien20}. 
We used CIGALE, as it models the SED of galaxies by conserving the energy balance between the dust-absorbed stellar emission and its re-emission in the IR.  
The capabilities of this SED fitting tool have already been verified on VIPERS observations \cite[i.e.,][Figueira et al., in prep.]{ralowski20,Vietri2021,turner21, pastis22}. 
Previous works have estimated stellar masses, SFRs, and AGN fractions of VIPERS' observed galaxies, and have shown consistency between each other. Moreover, we checked the stellar masses and absolute magnitudes obtained from CIGALE and Le Phare \citep{arnouts99, ilbert06} and found homogeneous results, which suggest that our SED fitting procedure is independent on the SED fitting methodology.
In this analysis, we assumed a delayed star formation history, \citet{bruzal03} single stellar population models with the initial mass function (IMF) given by \citet{chabrier03}, \citet{CF2000} attenuation law and the dust emission models of \citet{dale14} to build a grid of models. 

In the case of stellar mass estimation, the used attenuation curve plays a significant role \cite[on average leading to disparities of a factor of two,][]{malek18,buat19}.
For that reason we decided to test two very well know attenuation laws: modified attenuation laws of \cite{CF2000}, and \cite{Calzetti2000}.
As the VIPERS sample lacks reliable IR data we decided to use \cite{Calzetti2000} law.
Here, we want to stress that the stellar mass obtained with \cite{CF2000} recipe is systematically higher by $\sim$ 0.1 dex than \cite{Calzetti2000}.
A more detailed description can be found in the App.~\ref{app:attenuation}.
Moreover, we suspect that the amount of dust in galaxies we are looking for is rather low and therefore  we do not expect to find a huge difference between both methods{. Nevertheless, in our SED fitting approach, for the dust attenuation model, we used a wide range of parameters, allowing us to construct and fit templates similar to normal-star forming galaxies with a significant amount of dust.}
The final input parameters used in SED fitting with CIGALE are presented in Table~\ref{tab:cigale_input}.
A detailed description of each module can be found in \citet{malek18} and \citet{Boquien20}.

CIGALE estimates the physical properties of galaxies by evaluating a generated grid of models on data to minimise the likelihood distribution. 
The quality of the fit is expressed by the reduced value of the $\chi^2$ parameter\footnote{the $\chi^2$ divided by the number of data points}. 
The physical properties and their uncertainties are estimated as the likelihood-weighted means and standard deviations.
The mean goodness of the fit in \textit{UCMG candidates} sample, assuming input as shown in Tab.\ref{tab:cigale_input}, is $\chi^2  \simeq0.8$, and the mean uncertainties of stellar masses are at the level of 19\%.
Hereafter, every time the stellar masses of \textit{UCMG candidates} is mentioned, we refer to $M_{star}$ obtained by our SED fitting.

\begin{table*}[ht] 
\centering
\caption{Report of the compactness formulae and redshift ranges of sources presented in the literature.
The number of \textit{UCMG candidates} according to each criterion is given in the fourth column. 
The last column provides sample size of \textit{UCMG candidates}, which meet the compactness criterion and the mass completeness, above $\log(M_{star}/\rm{M}_{\odot}) \geq 10.86$ \citep{davidzon16}.
Two works \citep{cassata13, vanderwel14} applied two different criteria, and we refer to them as a \lq compact' and \lq ultracompact' for less and more restrictive, respectively.}
\begin{tabular}{ l c l c c }
    
Reference & Redshift& Formula & Number of sources & Mass complete\\
\hline
\cite{damjanov15}      & 0.24 - 0.66   & (log($R_e$) + 5.74)/log(M) < 0.568                & 4\,347 & 1\,664\\
\cite{cassata11} -- compact      & 1.20 - 3.00   & (log($R_e$) + 5.5)/log(M) < 0.54                  & 3\,139 & 1\,115\\
\cite{barro13}         & 1.40 - 3.00   & log(M/$R_e^{1.5}$) > 10.3                         & 3\,083 & 1\,370\\
\cite{vanderwel14} -- compact    & 0.00 - 3.00   & $R_e$/(M/10$^{11}$)$^{0.75}$ < 2.5 kpc            & 1\,801 &\phantom0\,914\\
\cite{charbonnier17}   & 0.20 - 0.60   & log(M) > 5 $\times$ 10$^{10}$, $R_e$ < 2 kpc      & 1\,061 &\phantom0\,372\\
\cite{spiniello21}     & 0.10 - 0.50   & log(M) > 6 $\times$ 10$^{10}$, $R_e$ < 2 kpc      & \phantom {2}\,693 &\phantom0\,372\\
\cite{buitrago18}      & 0.02 - 0.30   & log(M) > 8 $\times$ 10$^{10}$, $R_e$ < 2 kpc      & \phantom {2}\,277 &\phantom0\,277\\
\cite{cassata11} -- ultracompact      & 1.20 - 3.00   & (log($R_e$) + 5.8)/log(M) < 0.54                  & \phantom {2}\,250 & \phantom0\,\phantom082\\
\cite{vanderwel14} -- ultracompact    &0.00 - 3.00    & $R_e$/(M/10$^{11}$)$^{0.75}$ < 1.5 kpc            & \phantom {2}\,241  &\phantom0\,134\\
\cite{trujillo09}  & 0.00 - 0.20   & log(M) > 8 $\times$ 10$^{10}$, $R_e$ < 1.5 kpc    & \phantom {22}\,86 &\phantom0\,\phantom086\\
\hline
\end{tabular}
\label{tab:compactness_cuts}
\end{table*}

\section{Final selection of red nugget sample}\label{Sec:red_nuggets_selection}
In this section, we present the final criteria used to select the population of VIPERS red nuggets from our sample of 6\,961 \textit{UCMG candidates}.
In particular, we considered different definitions of compactness given by the limits in size, $R_e$, and stellar mass, $M_{star}$, followed by restricting the sample to red, passive galaxies based on their colours and star formation rate (hereafter SFR).

\subsection{Compactness}\label{sec:compactness}

The criterion used to select UCMGs, in particular the threshold for stellar mass and effective radius, has a great influence on the size and properties of the selected sample. 
Several different studies defined the class of massive compact galaxies based on various selection criteria \citep[e.g][]{trujillo09, damjanov15, charbonnier17}. 
To make a reliable comparison with literature, we adopt different definitions following selections that other authors have used.
The list of criteria applied to select massive and compact galaxies is given in Table~\ref{tab:compactness_cuts}. 
Different criteria can change the number of UCMGs by a factor of $\sim$50.
In particular, using the least conservative criterion proposed by \cite{damjanov15}, we end up with 4\,347 UCMGs, while the criterion of \cite{charbonnier17} defined in the same redshift range, results in 1\,061 galaxies. 
In our VIPERS red nuggets catalogue we included only sources, which meet one of the most restrictive criteria given by authors.
We used criterion proposed by \cite{trujillo09}:
$M_{star} > 8\times10^{10}$ ($\log(M_{star}/\rm{M}_{\odot}) \gtrsim 10.9$)
and $R_e > 1.5$~kpc, which limits the sample to only 86 objects (hereafter VIPERS UCMGs).

We performed a test, we found that if applying compactness and stellar mass completeness cuts together, the \cite{cassata11} criterion gives fewer sources above the stellar mass completeness threshold (82 red nuggets candidates  versus 86 from \cite{trujillo09} criterion).
However, we wanted to focus on truly compact objects, thus we used the \cite{trujillo09} criterion, which is very strict in the $R_e$.
Moreover \cite{trujillo09}  criterion is easiest to perform as it has two separates cuts for stellar mass and effective radius. 
I allows to control not only the compactness cuts, but also it is easy to compare with mass completeness.

The \cite{trujillo09} criterion was used in the literature to select UCMGs up to $z\sim0.5$ \citep{tortora16, scognamiglio20}. 
Although, this work aims to select UCMGs at the intermediate redshift range ($0.5<z<1.0$), we do not expect significant changes to the selection criterion. 
We assume that if UCMGs survive from $z > 3$ to the local Universe, $z < 1$, then their main physical properties such as $R_e$ and $M_{star}$ do not change significantly. This implies that they evolved relatively unaltered, without mergers or influence of other galaxies.
For that reason, we can use the same compactness criterion for UCMGs from $z \sim 2$ to $z \sim 0$.

According to \cite{davidzon16}, applying a cut in stellar mass assures stellar mass completeness in the VIPERS sample up to $z = 0.9$ ($\log(M_{star}/\rm{M}_{\odot}) \geq 10.86$).
We calculated the  stellar mass threshold  above which passive galaxies can be considered complete in stellar mass  in the redshift range 0.9--1.0. We have found, using the \cite{pozzetti10} method, which was also used in \cite{davidzon16}, the stellar mass threshold at the limit of ${\log}(M_{star}/\rm{M}_{\odot})=11.03$.
However, taking into account the low number of known red nuggets in the intermediate redshift, we decided to enlarge our analysis using the same cut in mass, 
${\log}(M_{star}/\rm{M}_{\odot})=10.86$, up to redshift 1, even with possible incompleteness bias for the most distant sources.
In this step of our selection, we also took into account the seeing-size limit.
As we mentioned before, the majority of \textit{UCMG candidates} (5\,673, $\sim$81\%) are larger than the size of pixel, but $\sim$73\% (63 galaxies) of VIPERS UCMGs have sizes measurements below seeing-defined limit (see Fig.~\ref{fig:Rz_pllt_with subsample}).
After visual inspection of a sample of 86 VIPERS UCMGs 
we found that images of all selected galaxies show compact and elliptical galaxy profiles with no signs of spiral arms or other morphological disturbances, and we decided to use all of them in our further analysis.

\subsection{Passiveness}\label{sec:passivnes}
To select red and passive galaxies, we performed multistage selection based on the colours, emission lines, and final visual inspection.

\subsubsection{NUVrK diagram}
\begin{figure}[ht]
\centering
\includegraphics[width = 0.49\textwidth]{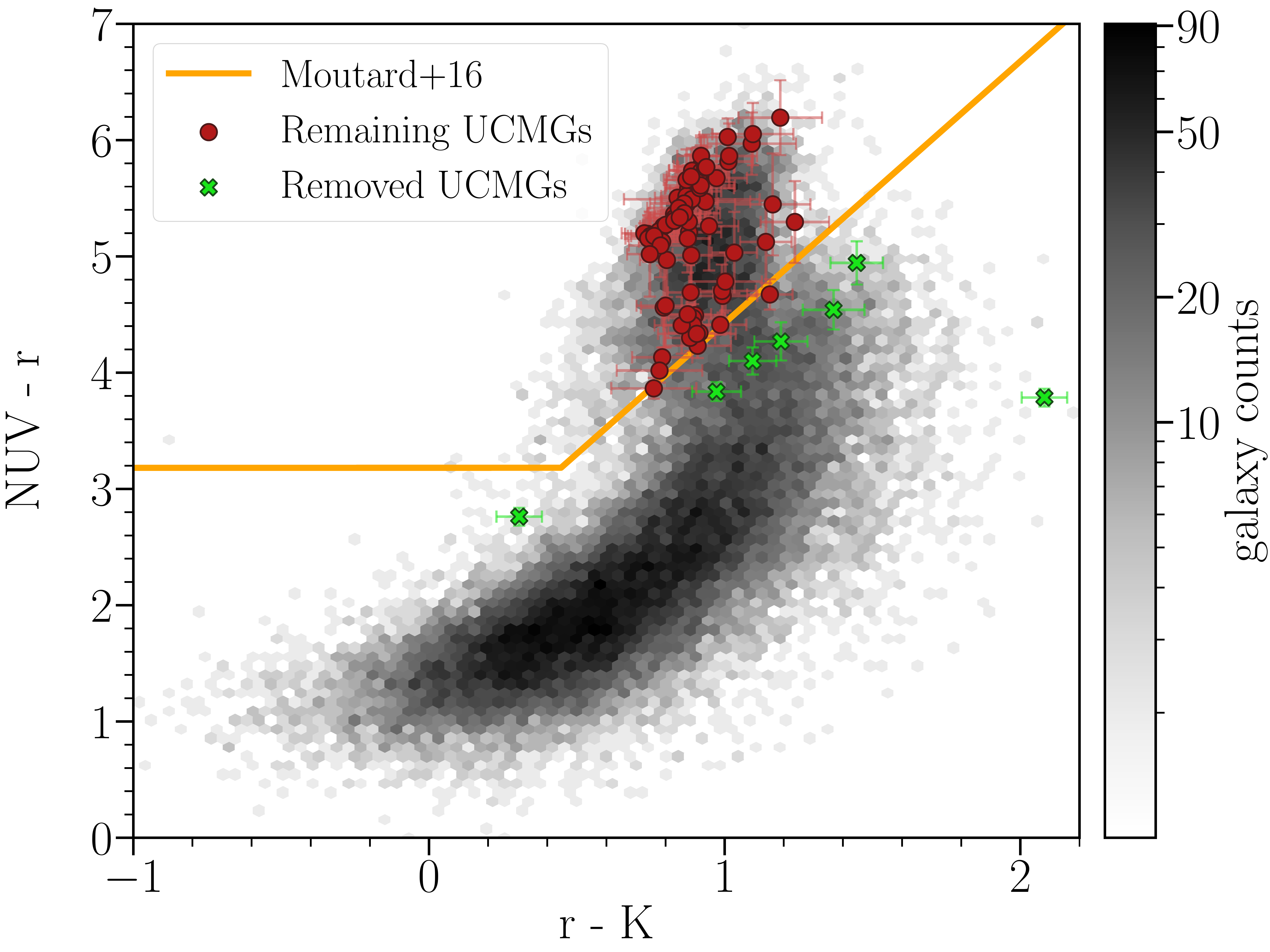}
\caption{NUVrK diagram. The distribution of 36\,157 galaxies (\textit{pure sample}) is shown in the background.
The orange line shows the limit for red galaxies adapted from \cite{moutard16_I}. The red and the green points represent our VIPERS UCMGs sample. Sources marked with green crosses are considered as blue (active) and have been removed from our sample.}
\label{fig:nuvrk}
\end{figure}

The primary criterion to select red sources used in our work is their position on the colour--colour diagram.
In our work, we used the NUVrK diagram, which is a tool to select a complete and pure sample of red passive galaxies, as this combination of colours allows to break dust-SFR degeneracy \citep{Arnouts2013AA}.
It is widely used by the VIPERS team, see for example:
\cite{fritz14},
\cite{davidzon16},
\cite{moutard16_I, moutard18},
\cite{gargiulo17},
\cite{siudek18_arx,siudek18},
\cite{turner21}.
The NUVrK diagram is shown in Fig~\ref{fig:nuvrk}.
All points (red and green) show the positions of 86 VIPERS UCMGs.
The orange line shows the boundary between red and blue galaxies derived by \cite{moutard16_I} obtained for the VIPERS survey.
As one can clearly see, seven galaxies lie below the limit, even while taking into account uncertainties.
Those seven galaxies are marked with green crosses and have been removed from our sample for the next steps (79 objects left).

\subsubsection{Visual inspection of VIPERS spectra}

\begin{figure}[ht]
\centering
\includegraphics[width = 0.49\textwidth]{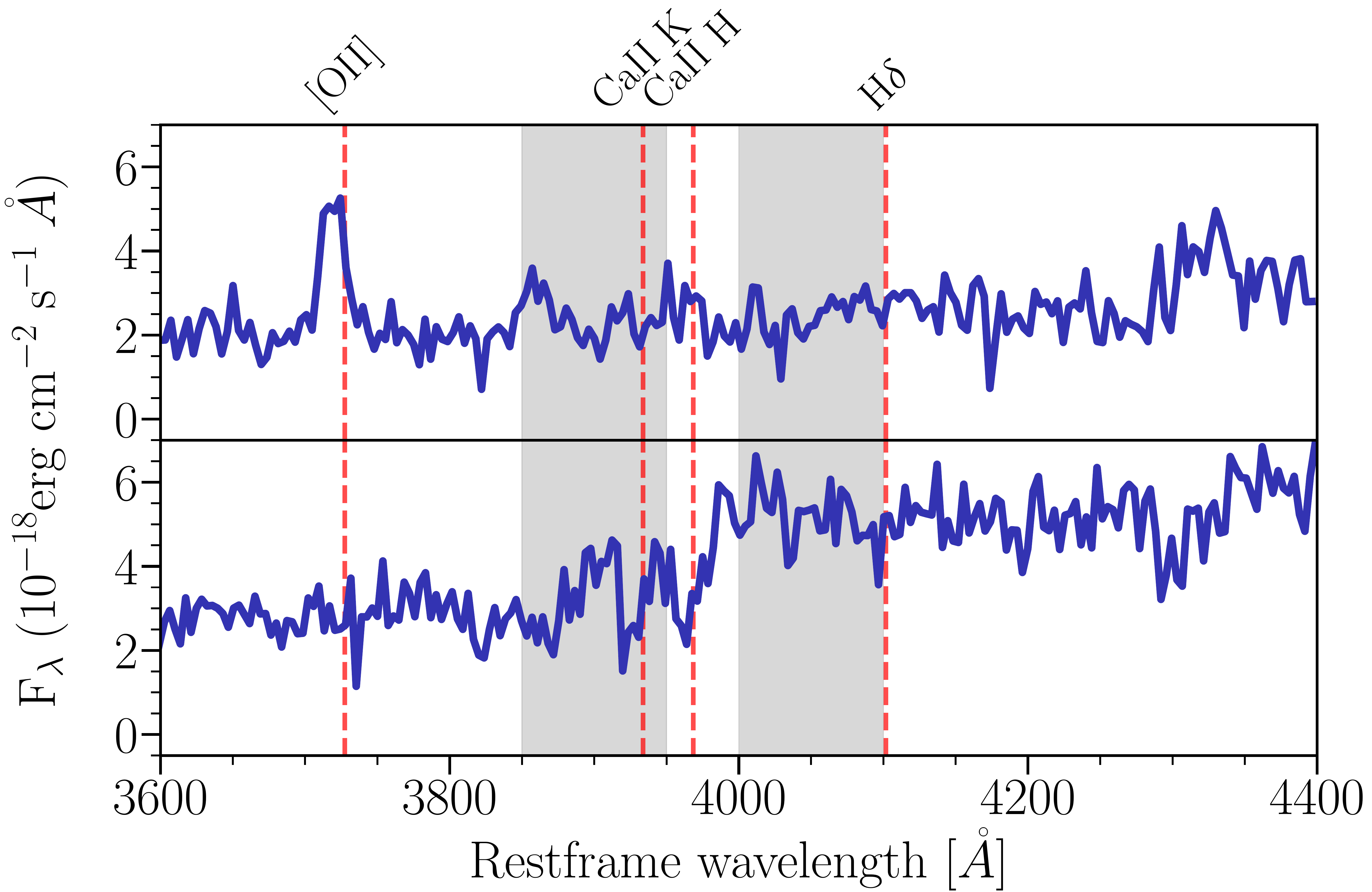}
\caption{Exemplary  spectra of galaxies.
The upper panel shows a possible star-forming galaxy while the bottom panel shows a likely passive galaxy.
With red dashed lines we marked from left: 
[OII] emision line(3727.5 \AA),
calcium K (3933.7 \AA),
calcium H (3968.5 \AA),
H$\delta$ (4101.7 \AA).
The grey shadings correspond to regions used for D4000 calculation \citep{Balogh1999}}
\label{fig:spectra_comparison}
\end{figure}

To verify the passiveness of the selected 79 red UCMGs, we checked their spectra obtained by VIPERS. 
We performed a detailed visual inspection of every object in the sample, searching for characteristic features which indicate star-forming activity, such as oxygen emission lines \citep{kennicutt92}, lack of CaII absorption lines or low Balmer break value \citep[hereafter D4000;][]{bruzal83, haines17}.
Figure~\ref{fig:spectra_comparison} shows examples of possibly active (top panel) and passive (bottom panel) galaxies in our sample of VIPERS UCMGs. 
The major differences are: the presence of a strong O[II] emission line, the lack of the CaII absorption lines and a weak D4000 in the active spectrum (top panel).
In our sample of 79 UCMGs, we found two objects which likely show active galaxy features. 
At this point, we left 77 UCMGs which we consider as red nugget candidates. 

\subsubsection{Sanity check}
\begin{figure}[ht]
\centering
\includegraphics[width = 0.49\textwidth]{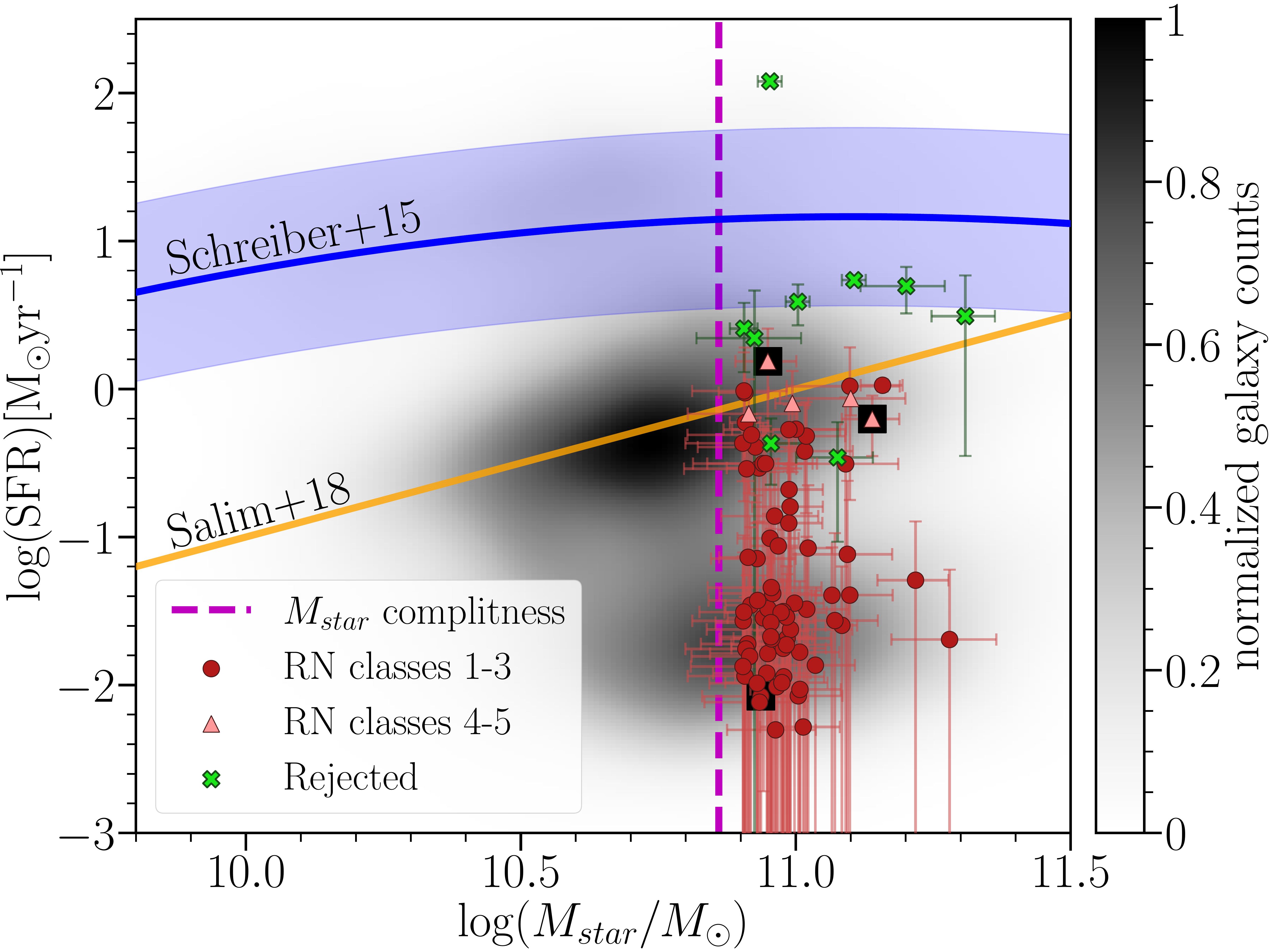}
\caption{Relation of SFR vs stellar mass. In the smoothed background the sample of 6\,961 \textit{UCMG candidates} is shown.
Points represent our red and green 86 UCMGs.
Red circles and pink triangles indicate VIPERS red nuggets (RN), while green crosses correspond to the UCMGs, which we considered as active. The orange line shows the limit for passive galaxies \citep{salim18}. The blue line shows the main sequence of galaxies according to \cite{schreiber15}. The magenta dashed line shows the mass completeness at redshift $z=$ 0.9 for passive galaxies equal to $\log(M_{star}/M_\odot) \sim 10.86$ according to \cite{davidzon16}. The three sources with relative errors higher than 45\% are marked with black squares.}
\label{fig:ssfr}
\end{figure}

As a  final sanity check of passiveness of the selected sample of 77 passive UCMGs,  we examined the stellar mass -- SFR relation, known also as the main sequence (hereafter MS).  
Figure \ref{fig:ssfr} shows the SFR as a function of the stellar mass of our 6\,961  galaxies which belong to the \textit{UCMG candidates}.
Red circles and pink triangles represent the sample of 77 VIPERS red nuggets candidates, and green crosses show the galaxies, which have been removed from our sample in previous steps.
The orange solid line shows the limit for passive objects used by \cite{salim18}, defined as specific SFR (SFR over stellar mass) equal to $10^{-11}$~$\rm{yr}^{-1}$.
The blue solid line shows the MS of star-forming galaxies derived by \cite{schreiber15}.
Here we plotted the MS at redshift $z \simeq 0.83$, which is the median redshift, of the sample of 86 VIPERS UCMGs.
In addition to that, the blue shaded region shows the range where $\log(\rm{SFR}/\rm{SFR}_{\rm{MS}}) < \log(4) \sim 0.6\rm{~dex}$.
It is a widely used limit to select both starbursting and passive galaxies from the main sequence relation \citep[eg:][]{elbaz18, buat19, donevski20}.
We found only three sources which were not removed in previous steps above the \cite{salim18} passiveness limit, but considering uncertainties in estimating both the SFR and stellar mass, as well the passive nature of their spectra, we decided to keep them in the sample. 
Taking into account the uncertainties, none of the 77 UCMGs is considered a MS galaxy.
For those reasons, we decided to not remove any additional sources.
Finally, we established a sample of confirmed VIPERS red nuggets containing 77 sources.

\section{Final catalogue of VIPERS red nuggets -  properties}\label{sec:properties}

In Table~\ref{tab:compactness_cuts} we listed the number of selected UCMGs corresponding to different compactness definitions from the literature. 
Within the VIPERS red nuggets catalogue we only considered the sample obtained using one of the most restrictive criteria for compactness,  with $M_{star} > 8\times10^{10}$ M$_\odot$ and $R_e<1.5$ kpc \citep{trujillo09}, resulting in 86 UCMGs. 
We then selected the 77 VIPERS red nuggets based on the NUVrK criterion, visual inspection of their spectra and additional sanity checks (see Sec.~\ref{sec:passivnes}).
Figure~\ref{fig:examples} shows four examples of selected red nuggets. 
We present not only the images in the \textit{u}, \textit{g}, \textit{r}, \textit{i}, and \textit{z} bands from the CFHT survey, but also the normalized spectra with marked oxygen [OII] (3727.5 \AA), calcium CaII~K (3933.7 \AA) and CaII~H (3968.5~\AA) lines.
The VIPERS red nuggets catalogue is publicly available and reported in Appendix~\ref{Sec:app_catalogue}.
In the catalogue, we list VIPERS IDs and positions in the sky, as well as all important parameters used in analysis such as: redshifts, effective radii, stellar masses and colours.

One characteristic that sets this study of red nuggets apart from other works is the unique redshift range over a wide-field systematic survey. 
In addition to that, our sample is already spectroscopically confirmed, which is an improvement in comparison with studies based only on photometry.
This means that we present the largest catalogue of spectroscopically identified red nuggets beyond the local Universe. 
In comparison, there is only 14 confirmed quiescent compact galaxies within the Baryon Oscillation Spectroscopic Survey (BOSS) at redshift $0.2 < z <0.6$ \citep{damjanov14} and about $\sim$20 found by \cite{charbonnier17} in the CFHT equatorial SDSS Stripe 82, also at redshift range $0.2<z<0.6$.

Within our final catalogue of red nuggets 96\% of all galaxies (74 sources) have relative errors of $R_e$ lower than 45\%. 
The remaining three red nuggets with the relative errors of $R_e$ equal to 99\%, 62\% and 61\%  included in our sample 
are highlighted in Table~\ref{Sec:app_catalogue} with underlined IDs and $R_e$ values, and in Fig.~\ref{fig:ssfr} with black squares.

\subsection{Comparison with VIPERS' results of unsupervised classification}

We performed a cross-match of our 77 red nuggets with detailed classification done by \cite{siudek18}.
In this paper, the authors classified 52\,114 VIPERS galaxies with the highest confidence ($>90\%$) of redshift measurements into 12 classes using the unsupervised machine learning algorithm, Fisher Expectation-Maximization~\citep[FEM;][]{bouveyron12}. 
All subclasses found by \cite{siudek18} mirror substructures in the bimodal colour distribution of galaxies, distinguishing subpopulations of passive (red), intermediate (green) and active (blue) galaxies and an additional class of broad-line AGNs. 
In the following section, the definition of the red, green and blue galaxy populations relies on FEM classification: 
\begin{itemize}
    \item red (subclasses 1–3), 
    \item green (subclasses 4–6),
    \item blue (subclasses 7–11).
\end{itemize}
The reliability of those three classes was checked in \cite{siudek18} via colour-colour method, spectral features like emission line distributions and the spectral continuum, and morphological parameters like S\'ersic index, stellar mass and SFR.
We refer to \cite{siudek18_arx, siudek18, siudek22} for a detailed description of the VIPERS classification. 
In this paper, we verify if VIPERS' red nuggets are preferably found in one of the red subclasses.

Our sample of 77 VIPERS red nuggets contains 72 red-class galaxies
(subclasses 1--3) marked in Fig.~\ref{fig:ssfr} as red circles, and five galaxies which belong to subclasses 4--5, green classes, represented in Fig.~\ref{fig:ssfr} with pink triangles.
All five green galaxies lie closer to the main sequence relation than the rest of the red nuggets, which additionally supports the reliability of FEM classification from \cite{siudek18}.
A histogram of red nuggets' classes is also presented in Fig.~\ref{fig:z_spec_hist}.
We did not find galaxies from the blue classes which supports our classification of passive, massive compact objects. 
Almost 65\% (49 out of 77) red nuggets are found in the subclass 1, which gathers massive, and  small red galaxies \citep[see Tab. 1 and Sec. 5.3 in][]{siudek22} 
This confirms the usefulness of applying unsupervised machine-learning algorithms to automatically find \lq peculiar\rq  galaxy subclasses \citep{siudek18_arx, siudek18, siudek22}.

\subsection{The D4000 distribution of intermediate-$z$ red nuggets}

The strength of the 4000~\AA ~spectral break \citep[hereafter D4000, defined as in][]{Balogh1999} is one of the main and direct characteristics of the star formation history of galaxies.
Its correlation with young and evolved stellar populations makes this spectral index suitable as a passiveness indicator \cite[i.e.][]{kauffmann03,Kauffmann2003b,Siudek2017, haines17}.

\begin{figure}[ht]
\centering
\includegraphics[width = 0.49\textwidth]{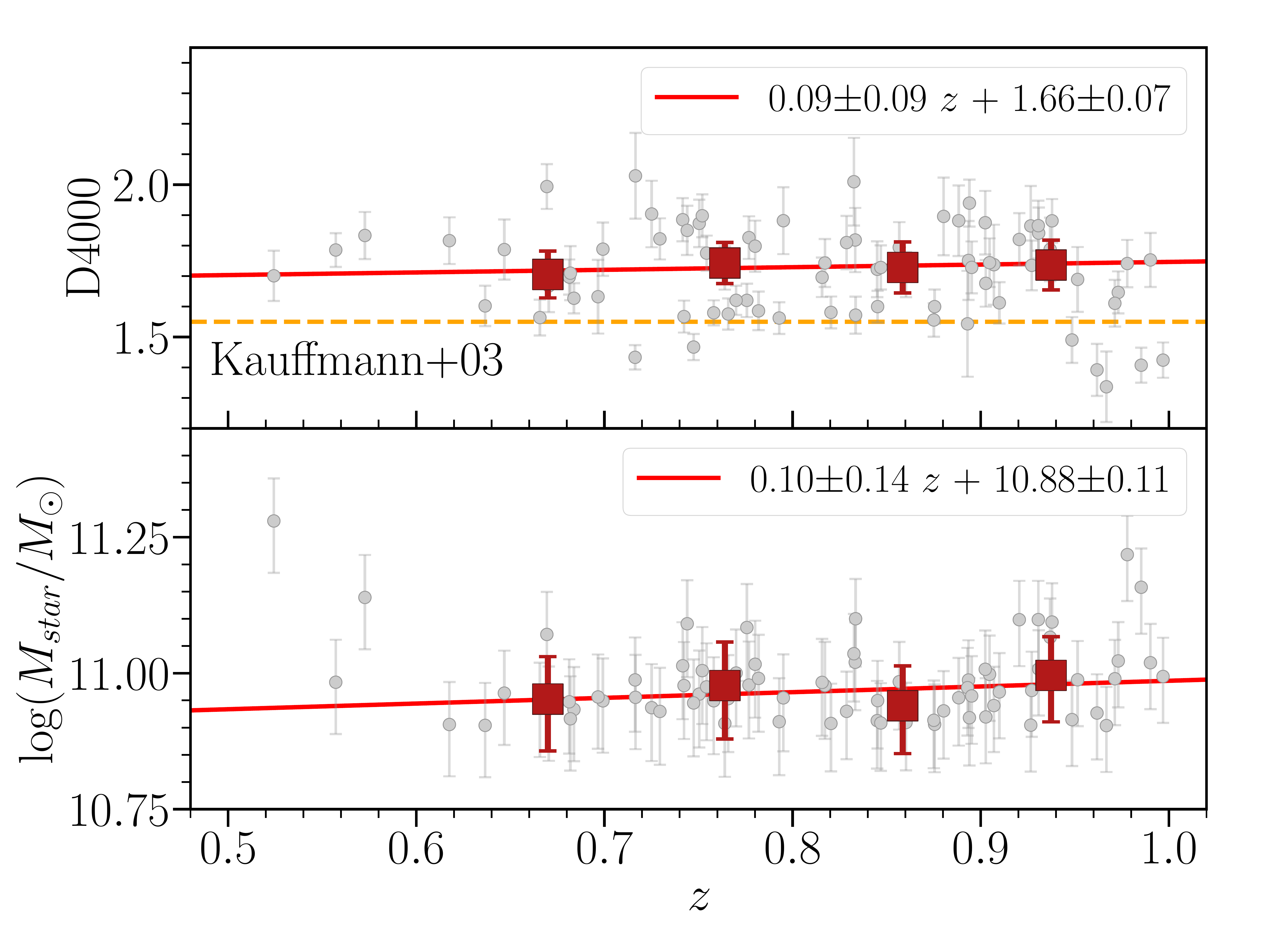}
\caption{Relation of D4000 vs $z$ in the top panel and $M_{star}$ vs $z$ in the bottom panel.
Grey points represent 77 red nuggets and the
red squares represent the median value in every redshift bin (see Tab.~\ref{tab:number_density}), and the red line is the best linear fit.
The dashed orange line represents D4000 equals 1.55, which is considered a passiveness limit derived by \cite{kauffmann03} in the local Universe.}
\label{fig:z_d4000}
\end{figure}

Figure \ref{fig:z_d4000} shows D4000 -- $z$ (top panel) and $M_{star}$  -- $z$ (bottom panel) relations, where grey points represent 77 red nuggets, the red squares are median values calculated for four redshift bins defined for similar number of objects (see Tab.~\ref{tab:number_density}). 
The red lines show the linear fit to the median values.
In addition to that, the orange dashed line indicates D4000 = 1.55 -- the passiveness boundary found by \cite{kauffmann03} for the local Universe based on the SDSS survey.
Taking into account uncertainties of D4000 measurements, six galaxies lie below the \cite{kauffmann03} limit.
Two of those galaxies have $z\sim0.73$, while four are located at the end of our redshift range, at $z>0.85$, where we can expect to see a younger population than in the local Universe.

Considering the median values in redshift bins, there is no significant D4000--$z$ dependence for the presented sample, and the median value of D4000 for red nuggets in the redshift range 0.5-1.0 stays constant at the level 1.66$\pm$0.05 indicating passiveness of the sample.
The bottom panel of Fig.~\ref{fig:z_d4000} shows the $M_{star}$--$z$ relation for the sample of 77 red nuggets. 
We found no clear dependence of the stellar mass on redshift.
However, the highest redshift bin may be biased by non negligible incompleteness.

\section{Discussion and results}\label{sec:discussion}

\subsection{Influence of compactness criteria on sample size}

All compactness criteria found in literature are functions of $R_e$ and $M_{star}$.
The discrepancy of those criteria is shown in Tab.~\ref{tab:compactness_cuts}. 
Unfortunately, we can not compare samples sizes of UCMGs found by different criteria, as not all of them are mass complete.
For this reason, we additionally limit the criteria by cutting all galaxies with stellar masses below the completeness limit.
In the VIPERS survey, the number of sources considered as compact and complete in stellar mass found here varies between 1\,664, while using the \cite{damjanov15} criterion, and 82 with \cite{cassata11} ultracompact criterion. 
We thus see that the choice of compactness criterion can influence the sample size by up to a factor of $\sim$20.

\subsection{Evolution of number densities}

\begin{figure}[ht]
\centering
\includegraphics[width = 0.49\textwidth]{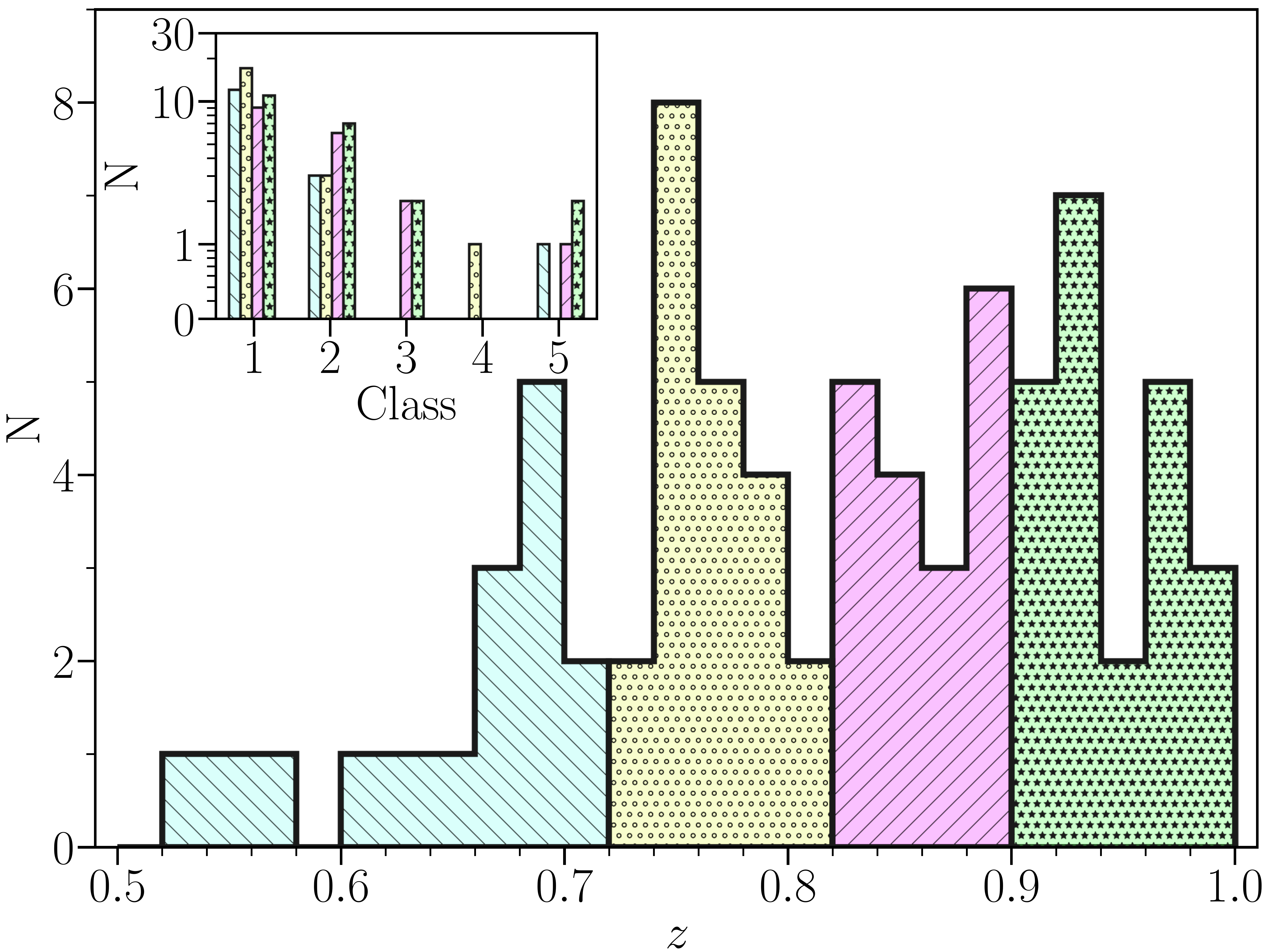}
\caption{Distribution of VIPERS red nuggets in redshift (big histogram) and in FEM classes (smaller one). Different colours correspond to our bins.}
\label{fig:z_spec_hist}
\end{figure}

\begin{figure}[ht]
\centering
\includegraphics[width = 0.49\textwidth]{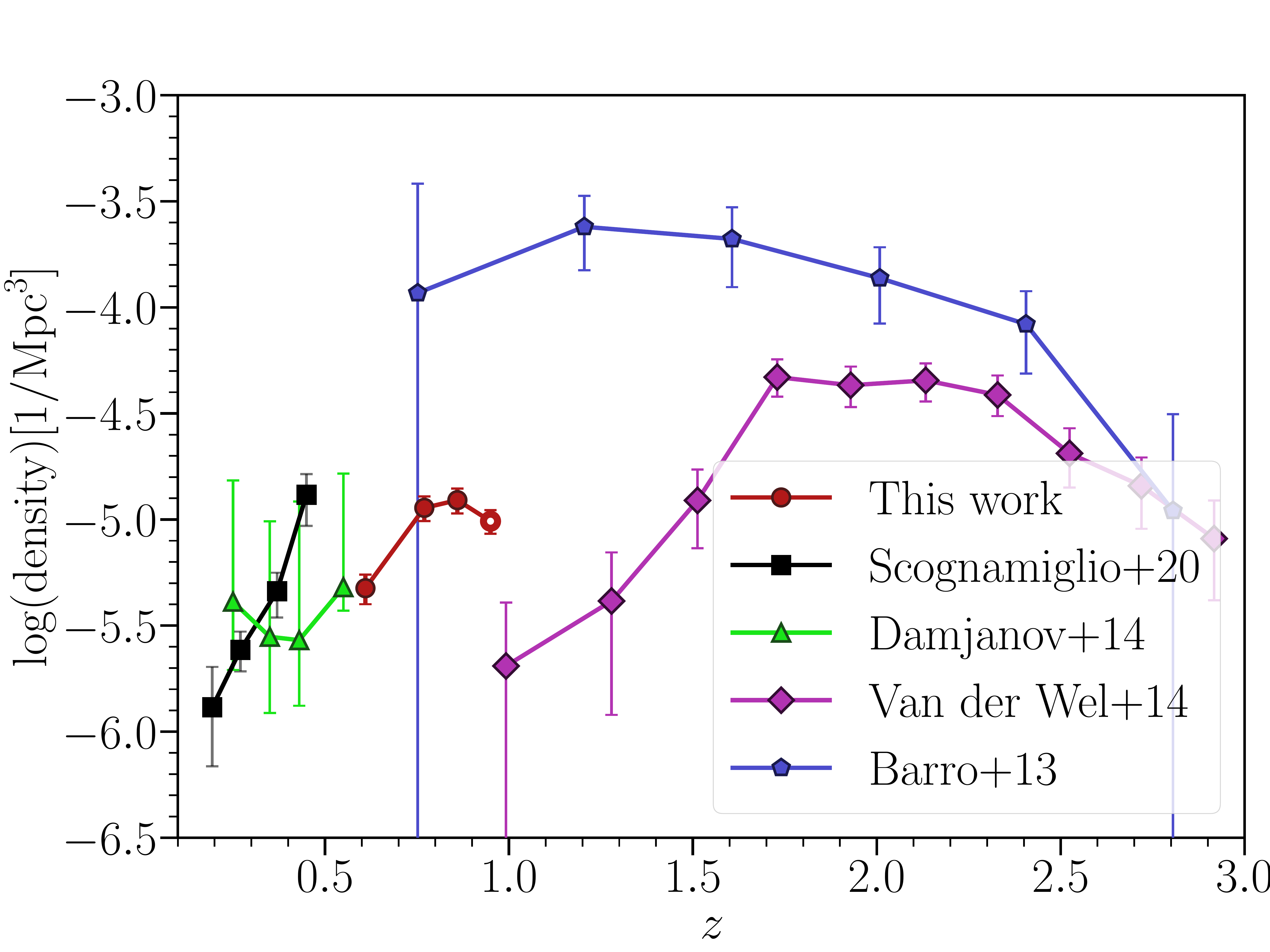}
\caption{Number density per comoving cubic Mpc vs redshift. The red points show results from this work, the green ones show the quiescent compact galaxies found by \cite{damjanov14}, the black ones show the compact galaxies found by \cite{scognamiglio20}, the violet ones show ultracompact ETGs analysed by \cite{vanderwel14} and the blue ones present the quiescent compact galaxies found by \cite{barro13}.}
\label{fig:literature_number_density}
\end{figure}

In order to derive number densities, we divided all red nuggets into four redshift bins containing a roughly equal number of galaxies per bin: 0.50--0.72, 0.72--0.82, 0.82--0.9 and 0.90--1.00  (see Table~\ref{tab:number_density}).
Taking into account that our sources are not evenly distributed in redshift (see Fig. \ref{fig:z_spec_hist}) our bins are not uniform.
We then calculated weighted number of sources based on TSR and SSR of the parent sample: 
\begin{equation}
    \rm{N_w} = \sum_{i}^{\rm{N}} (\rm{TSR}_i \cdot\rm{SSR}_i)^{-1},
\end{equation}
\citep[for details see][]{garilli14}.
Finally, we normalized our bins to the comoving volume corresponding to the VIPERS observed area and calculated the number density per cubic comoving Mpc.
The results are presented in Table \ref{tab:number_density} and in Figure~\ref{fig:literature_number_density}.

\begin{table}[ht]
\centering
	\caption{Overview of the four redshift bins and the summary of all four bins. The second column shows the number of sources in each bin, and the third one presents the weighted number of sources, which takes into account the TSR and SSR. In the last column, we show the number density per comoving cubic Mpc.}
	\label{tab:number_density}
	\footnotesize
	\begin{tabular}{c c c c}
\hline
Redshift range & N & N$_{\rm{w}}$& Number density [$\rm{Mpc}^{-3}$] \\
\hline
$0.50  \leq z \leq 0.72$  & 16 & 39.25 & $4.74 \times10^{-6}$\\
$0.72 < z \leq 0.82 $ &21 & 56.75 & $1.14 \times10^{-5}$\\
$0.82 < z \leq 0.90 $ &18 & 55.64 & $1.23 \times10^{-5}$\\
\,\,$0.90 < z \leq 1.00 $\tablefootnote{Lower limit.} &22 & 61.05 & $9.82 \times10^{-6}$\\

\hline
$0.50 \leq z \leq 1.00$ & 77 & 212.69 & $8.86 \times 10^{-6}$\\
\hline
\end{tabular}
\end{table}

Figure \ref{fig:literature_number_density} shows number densities (see Tab.~\ref{tab:number_density}) as a function of redshift.
Number densities of VIPERS red nuggets calculated in the first three redshift bins are marked with the full redcircles, while the results from the redshift bin 0.9--1.0 which can be biased by the mass incompleteness, is shown as a red open circle.
We stressed that due to the mass incompleteness presented number density for this redshift bin should be considered as a lower limit.
Presented uncertainties of the number counts take into account fluctuations due to the Poisson noise.

We compared our results with the number densities of quiescent compact galaxies found by \cite{damjanov14} in the Baryon Oscillation Spectroscopic Survey, which are marked with green triangles.
In addition to that, we also plotted the number densities of ultracompact massive galaxies found in the Kilo Degree Survey by \cite{scognamiglio20}. 
Those results are shown as black squares.
As we do not expect to find in this sample a statistically significant number of star-forming massive compact objects (for example in our initial sample of 86 UCMGs only $\sim$10\% were considered as not passive), it is reasonable to compare those two groups.
At higher redshifts, we present passive ultracompact galaxies found in the CANDELS field by \cite{vanderwel14} marked with purple diamonds (found with the ultracompact criterion, see Tab. \ref{tab:compactness_cuts}).
Moreover, we also show the quiescent compact sources found in the GOODS-S and UDS fields of CANDELS by \cite{barro13} with blue pentagons.
None of the high-redshift studies provide explicit number densities, and an online tool was used to extract their corresponding values directly from plots\footnote{the tool which we used to extract values of the number densities is available here: www.graphreader.com}.

Figure~\ref{fig:literature_number_density} shows that our results are in overall agreement with the lower-redshift studies, to within an order of magnitude.
However, the trends found by \cite{scognamiglio20} and \cite{damjanov14} are slightly different, as those samples were selected using a slightly different approach:
(1) \cite{damjanov14} selected objects with  effective radius smaller than $R_e < 2$~kpc (in our work $R_e < 1.5$ kpc) while the cut on stellar mass is the same as the one used in this work ($M_{star} < 8\times10^{10}$).
Moreover, the resulting numbers from \cite{damjanov14} are considered as the lower limits for number density due to sample incompleteness.
(2) \cite{scognamiglio20} performed a statistical correction, taking into account false positives and false negatives.
For the first three bins (redshift lower than 0.4), this correction does not change the number density values significantly (uncorrected results are still inside error bars).
However, the uncorrected value of the last bin is lower by around twice the error value. 
In Figure \ref{fig:literature_number_density} we show only corrected number densities. 
The criteria used for selecting compact sources are exactly the same as the ones we used ($M_{star} < 8\times10^{10}$ and $R_e < 1.5$~kpc).
However, the authors do not discuss the subject of passiveness.

To summarize, Figure \ref{fig:literature_number_density} shows that, our results match the ones presented in \cite{damjanov14}.
Similarly, the results obtained by \cite{scognamiglio20}, except the number density at $z\sim$0.45, match smoothly with our results.
However, as mentioned before, the uncorrected value of the last point is $\sim$0.3 dex lower.

On the other hand, at high-redshifts, the differences are significant.
However, both \cite{vanderwel14} and \cite{barro13} used radically different, less conservative, compactness criteria from those used in this work.
In addition to that, we cannot straightforwardly compare the number of VIPERS UCMGs selected using the same criteria due to the mass completeness.
For this reason, here only the mass complete sample sizes are discussed (see Tab.~\ref{tab:compactness_cuts}). 
Considering only mass complete samples of UCMGs, we find that the \cite{trujillo09} criterion (our fiducial choice throughout this work) yields roughly 1.5 and 14 times fewer compact sources than the \cite{barro13} and \cite{vanderwel14} criteria, respectively.
Comparing the number density presented by \cite{barro13} at $z\sim$0.75 to the number density of the VIPERS red nuggets at $z\sim$0.63, we find the former to be $\sim$20 times larger.
As the ratio of number densities is similar to our inferred ratio of the number of sources found with the different criteria, we conclude that our results are consistent with those of \cite{barro13}.
Nevertheless, the number density of VIPERS red nuggets at $z\sim$0.95  is roughly 5 times larger than the number density reported by  \cite{vanderwel14} at z$\sim$1. The cause of this disagreement is unclear.

\section{Summary}\label{Sec:summary}
We have shown that the sample size of UCMGs in the uniform spectroscopic VIPERS survey is strongly dependent on the compactness criterion, and comparing the different criteria is not straightforward.
Our catalogue of 77 VIPERS red nuggets significantly increases the number of known sources.
The brand new catalogue of red nuggets presented here is a solid framework for future studies.
Very restricted selection cut (based on $R_e$, $M_{star}$ and  passiveness indicators) provides a pure sample of intermediate-redshift red nuggets.
The evolution of the number density of red nuggets and relics over cosmic time is still an open issue.
The results presented here provide a next step to understanding the evolution of red nuggets. 
We were able to trace the number densities of red nuggets with a comparable statistical significance to the one found at lower redshifts \citep{charbonnier17,scognamiglio20}.
The detailed analysis of their physical properties to reveal their nature is left for future work and their environmental properties are discussed in Siudek et al. in prep.

\begin{acknowledgements} 
We would like to thank the anonymous referee for the careful reading of the manuscript and the very useful comments.
We thank Diana Scognamiglio for providing data for our analysis and Bianca Garilli for useful comments and discussions.
KL and KM  are grateful for support from the Polish National Science Centre via grant UMO-2018/30/E/ST9/00082. 
AK acknowledges support from the First TEAM grant of the Foundation for Polish Science No. POIR.04.04.00-00-5D21/18-00 and the Polish National Agency for Academic Exchange grant No. BPN/BEK/2021/1/00319/DEC/1.
MS has been supported by the Polish National Agency for Academic Exchange (Bekker grant BPN/BEK/2021/1/00298/DEC/1), the European Union's Horizon 2020 Research and Innovation Programme under the Maria Sklodowska-Curie grant agreement (No. 754510),  the National Science Centre of Poland (grant UMO-2016/23/N/ST9/02963) and the Spanish Ministry of Science and Innovation through the Juan de la Cierva-formacion programme (FJC2018-038792-I). 
This work has been supported by the Polish National Science Centre (UMO-2018/30/M/ST9/00757), and by Polish Ministry of Science and Higher Education grant DIR/WK/2018/12.
\end{acknowledgements}

\bibliographystyle{aa}
\bibliography{main}

\begin{thebibliography}{85}
\expandafter\ifx\csname natexlab\endcsname\relax\def\natexlab#1{#1}\fi

\bibitem[{{Ahumada} {et~al.}(2020){Ahumada}, {Prieto}, {Almeida}, {Anders},
  {Anderson}, {Andrews}, {Anguiano}, {Arcodia}, {Armengaud}, {Aubert}, {Avila},
  {Avila-Reese}, {Badenes}, {Balland}, {Barger}, {Barrera-Ballesteros}, {Basu},
  {Bautista}, {Beaton}, {Beers}, {Benavides}, {Bender}, {Bernardi}, {Bershady},
  {Beutler}, {Bidin}, {Bird}, {Bizyaev}, {Blanc}, {Blanton}, {Boquien},
  {Borissova}, {Bovy}, {Brandt}, {Brinkmann}, {Brownstein}, {Bundy}, {Bureau},
  {Burgasser}, {Burtin}, {Cano-D{\'\i}az}, {Capasso}, {Cappellari}, {Carrera},
  {Chabanier}, {Chaplin}, {Chapman}, {Cherinka}, {Chiappini}, {Doohyun Choi},
  {Chojnowski}, {Chung}, {Clerc}, {Coffey}, {Comerford}, {Comparat}, {da
  Costa}, {Cousinou}, {Covey}, {Crane}, {Cunha}, {Ilha}, {Dai}, {Damsted},
  {Darling}, {Davidson}, {Davies}, {Dawson}, {De}, {de la Macorra}, {De Lee},
  {Queiroz}, {Deconto Machado}, {de la Torre}, {Dell'Agli}, {du Mas des
  Bourboux}, {Diamond-Stanic}, {Dillon}, {Donor}, {Drory}, {Duckworth},
  {Dwelly}, {Ebelke}, {Eftekharzadeh}, {Davis Eigenbrot}, {Elsworth},
  {Eracleous}, {Erfanianfar}, {Escoffier}, {Fan}, {Farr},
  {Fern{\'a}ndez-Trincado}, {Feuillet}, {Finoguenov}, {Fofie},
  {Fraser-McKelvie}, {Frinchaboy}, {Fromenteau}, {Fu}, {Galbany}, {Garcia},
  {Garc{\'\i}a-Hern{\'a}ndez}, {Oehmichen}, {Ge}, {Maia}, {Geisler}, {Gelfand},
  {Goddy}, {Gonzalez-Perez}, {Grabowski}, {Green}, {Grier}, {Guo}, {Guy},
  {Harding}, {Hasselquist}, {Hawken}, {Hayes}, {Hearty}, {Hekker}, {Hogg},
  {Holtzman}, {Horta}, {Hou}, {Hsieh}, {Huber}, {Hunt}, {Chitham}, {Imig},
  {Jaber}, {Angel}, {Johnson}, {Jones}, {J{\"o}nsson}, {Jullo}, {Kim},
  {Kinemuchi}, {Kirkpatrick}, {Kite}, {Klaene}, {Kneib}, {Kollmeier}, {Kong},
  {Kounkel}, {Krishnarao}, {Lacerna}, {Lan}, {Lane}, {Law}, {Le Goff}, {Leung},
  {Lewis}, {Li}, {Lian}, {Lin}, {Long}, {Longa-Pe{\~n}a}, {Lundgren}, {Lyke},
  {Ted Mackereth}, {MacLeod}, {Majewski}, {Manchado}, {Maraston}, {Martini},
  {Masseron}, {Masters}, {Mathur}, {McDermid}, {Merloni}, {Merrifield},
  {M{\'e}sz{\'a}ros}, {Miglio}, {Minniti}, {Minsley}, {Miyaji}, {Mohammad},
  {Mosser}, {Mueller}, {Muna}, {Mu{\~n}oz-Guti{\'e}rrez}, {Myers}, {Nadathur},
  {Nair}, {Nandra}, {do Nascimento}, {Nevin}, {Newman}, {Nidever}, {Nitschelm},
  {Noterdaeme}, {O'Connell}, {Olmstead}, {Oravetz}, {Oravetz}, {Osorio},
  {Pace}, {Padilla}, {Palanque-Delabrouille}, {Palicio}, {Pan}, {Pan},
  {Parker}, {Paviot}, {Peirani}, {Ram{\'r}ez}, {Penny}, {Percival},
  {Perez-Fournon}, {P{\'e}rez-R{\`a}fols}, {Petitjean}, {Pieri},
  {Pinsonneault}, {Poovelil}, {Povick}, {Prakash}, {Price-Whelan}, {Raddick},
  {Raichoor}, {Ray}, {Rembold}, {Rezaie}, {Riffel}, {Riffel}, {Rix}, {Robin},
  {Roman-Lopes}, {Rom{\'a}n-Z{\'u}{\~n}iga}, {Rose}, {Ross}, {Rossi},
  {Rowlands}, {Rubin}, {Salvato}, {S{\'a}nchez}, {S{\'a}nchez-Menguiano},
  {S{\'a}nchez-Gallego}, {Sayres}, {Schaefer}, {Schiavon}, {Schimoia},
  {Schlafly}, {Schlegel}, {Schneider}, {Schultheis}, {Schwope}, {Seo},
  {Serenelli}, {Shafieloo}, {Shamsi}, {Shao}, {Shen}, {Shetrone}, {Shirley},
  {Aguirre}, {Simon}, {Skrutskie}, {Slosar}, {Smethurst}, {Sobeck}, {Sodi},
  {Souto}, {Stark}, {Stassun}, {Steinmetz}, {Stello}, {Stermer},
  {Storchi-Bergmann}, {Streblyanska}, {Stringfellow}, {Stutz}, {Su{\'a}rez},
  {Sun}, {Taghizadeh-Popp}, {Talbot}, {Tayar}, {Thakar}, {Theriault}, {Thomas},
  {Thomas}, {Tinker}, {Tojeiro}, {Toledo}, {Tremonti}, {Troup}, {Tuttle},
  {Unda-Sanzana}, {Valentini}, {Vargas-Gonz{\'a}lez}, {Vargas-Maga{\~n}a},
  {V{\'a}zquez-Mata}, {Vivek}, {Wake}, {Wang}, {Weaver}, {Weijmans}, {Wild},
  {Wilson}, {Wilson}, {Wolthuis}, {Wood-Vasey}, {Yan}, {Yang}, {Y{\`e}che},
  {Zamora}, {Zarrouk}, {Zasowski}, {Zhang}, {Zhao}, {Zhao}, {Zheng}, {Zheng},
  {Zhu}, \& {Zou}}]{ahmuda20}
{Ahumada}, R., {Prieto}, C.~A., {Almeida}, A., {et~al.} 2020, \apjs, 249, 3

\bibitem[{{Arnouts} {et~al.}(1999){Arnouts}, {Cristiani}, {Moscardini},
  {Matarrese}, {Lucchin}, {Fontana}, \& {Giallongo}}]{arnouts99}
{Arnouts}, S., {Cristiani}, S., {Moscardini}, L., {et~al.} 1999, \mnras, 310,
  540

\bibitem[{{Arnouts} {et~al.}(2013){Arnouts}, {Le Floc'h}, {Chevallard},
  {Johnson}, {Ilbert}, {Treyer}, {Aussel}, {Capak}, {Sanders}, {Scoville},
  {McCracken}, {Milliard}, {Pozzetti}, \& {Salvato}}]{Arnouts2013AA}
{Arnouts}, S., {Le Floc'h}, E., {Chevallard}, J., {et~al.} 2013, \aap, 558, A67

\bibitem[{{Balogh} {et~al.}(1999){Balogh}, {Morris}, {Yee}, {Carlberg}, \&
  {Ellingson}}]{Balogh1999}
{Balogh}, M.~L., {Morris}, S.~L., {Yee}, H.~K.~C., {Carlberg}, R.~G., \&
  {Ellingson}, E. 1999, \apj, 527, 54

\bibitem[{{Barro} {et~al.}(2013){Barro}, {Faber}, {P{\'e}rez-Gonz{\'a}lez},
  {Koo}, {Williams}, {Kocevski}, {Trump}, {Mozena}, {McGrath}, {van der Wel},
  {Wuyts}, {Bell}, {Croton}, {Ceverino}, {Dekel}, {Ashby}, {Cheung},
  {Ferguson}, {Fontana}, {Fang}, {Giavalisco}, {Grogin}, {Guo}, {Hathi},
  {Hopkins}, {Huang}, {Koekemoer}, {Kartaltepe}, {Lee}, {Newman}, {Porter},
  {Primack}, {Ryan}, {Rosario}, {Somerville}, {Salvato}, \& {Hsu}}]{barro13}
{Barro}, G., {Faber}, S.~M., {P{\'e}rez-Gonz{\'a}lez}, P.~G., {et~al.} 2013,
  \apj, 765, 104

\bibitem[{{Boquien} {et~al.}(2019){Boquien}, {Burgarella}, {Roehlly}, {Buat},
  {Ciesla}, {Corre}, {Inoue}, \& {Salas}}]{Boquien20}
{Boquien}, M., {Burgarella}, D., {Roehlly}, Y., {et~al.} 2019, \aap, 622, A103

\bibitem[{{Bouveyron} \& {Brunet}(2012)}]{bouveyron12}
{Bouveyron}, C. \& {Brunet}, C. 2012, arXiv e-prints, arXiv:1204.2067

\bibitem[{{Bruzual} \& {Charlot}(2003)}]{bruzal03}
{Bruzual}, G. \& {Charlot}, S. 2003, \mnras, 344, 1000

\bibitem[{{Bruzual A.}(1983)}]{bruzal83}
{Bruzual A.}, G. 1983, \apj, 273, 105

\bibitem[{{Buat} {et~al.}(2019){Buat}, {Ciesla}, {Boquien}, {Ma{\l}ek}, \&
  {Burgarella}}]{buat19}
{Buat}, V., {Ciesla}, L., {Boquien}, M., {Ma{\l}ek}, K., \& {Burgarella}, D.
  2019, \aap, 632, A79

\bibitem[{{Buitrago} {et~al.}(2018){Buitrago}, {Ferreras}, {Kelvin}, {Baldry},
  {Davies}, {Angthopo}, {Khochfar}, {Hopkins}, {Driver}, {Brough}, {Sabater},
  {Conselice}, {Liske}, {Holwerda}, {Bremer}, {Phillipps},
  {L{\'o}pez-S{\'a}nchez}, \& {Graham}}]{buitrago18}
{Buitrago}, F., {Ferreras}, I., {Kelvin}, L.~S., {et~al.} 2018, \aap, 619, A137

\bibitem[{{Calzetti} {et~al.}(2000){Calzetti}, {Armus}, {Bohlin}, {Kinney},
  {Koornneef}, \& {Storchi-Bergmann}}]{Calzetti2000}
{Calzetti}, D., {Armus}, L., {Bohlin}, R.~C., {et~al.} 2000, \apj, 533, 682

\bibitem[{{Cassata} {et~al.}(2011){Cassata}, {Giavalisco}, {Guo}, {Renzini},
  {Ferguson}, {Koekemoer}, {Salimbeni}, {Scarlata}, {Grogin}, {Conselice},
  {Dahlen}, {Lotz}, {Dickinson}, \& {Lin}}]{cassata11}
{Cassata}, P., {Giavalisco}, M., {Guo}, Y., {et~al.} 2011, \apj, 743, 96

\bibitem[{{Cassata} {et~al.}(2013){Cassata}, {Giavalisco}, {Williams}, {Guo},
  {Lee}, {Renzini}, {Ferguson}, {Faber}, {Barro}, {McIntosh}, {Lu}, {Bell},
  {Koo}, {Papovich}, {Ryan}, {Conselice}, {Grogin}, {Koekemoer}, \&
  {Hathi}}]{cassata13}
{Cassata}, P., {Giavalisco}, M., {Williams}, C.~C., {et~al.} 2013, \apj, 775,
  106

\bibitem[{{Chabrier}(2003)}]{chabrier03}
{Chabrier}, G. 2003, \pasp, 115, 763

\bibitem[{{Charbonnier} {et~al.}(2017){Charbonnier}, {Huertas-Company},
  {Gon{\c{c}}alves}, {Men{\'e}ndez-Delmestre}, {Bundy}, {Galliano}, {Moraes},
  {Makler}, {Pereira}, {Erben}, {Hildebrandt}, {Shan}, {Caminha}, {Grossi}, \&
  {Riguccini}}]{charbonnier17}
{Charbonnier}, A., {Huertas-Company}, M., {Gon{\c{c}}alves}, T.~S., {et~al.}
  2017, \mnras, 469, 4523

\bibitem[{{Charlot} \& {Fall}(2000)}]{CF2000}
{Charlot}, S. \& {Fall}, S.~M. 2000, \apj, 539, 718

\bibitem[{{Colless} {et~al.}(2001){Colless}, {Dalton}, {Maddox}, {Sutherland},
  {Norberg}, {Cole}, {Bland-Hawthorn}, {Bridges}, {Cannon}, {Collins}, {Couch},
  {Cross}, {Deeley}, {De Propris}, {Driver}, {Efstathiou}, {Ellis}, {Frenk},
  {Glazebrook}, {Jackson}, {Lahav}, {Lewis}, {Lumsden}, {Madgwick}, {Peacock},
  {Peterson}, {Price}, {Seaborne}, \& {Taylor}}]{colless01}
{Colless}, M., {Dalton}, G., {Maddox}, S., {et~al.} 2001, \mnras, 328, 1039

\bibitem[{{Daddi} {et~al.}(2005){Daddi}, {Renzini}, {Pirzkal}, {Cimatti},
  {Malhotra}, {Stiavelli}, {Xu}, {Pasquali}, {Rhoads}, {Brusa}, {di Serego
  Alighieri}, {Ferguson}, {Koekemoer}, {Moustakas}, {Panagia}, \&
  {Windhorst}}]{daddi2005}
{Daddi}, E., {Renzini}, A., {Pirzkal}, N., {et~al.} 2005, \apj, 626, 680

\bibitem[{{Dale} {et~al.}(2014){Dale}, {Helou}, {Magdis}, {Armus},
  {D{\'\i}az-Santos}, \& {Shi}}]{dale14}
{Dale}, D.~A., {Helou}, G., {Magdis}, G.~E., {et~al.} 2014, \apj, 784, 83

\bibitem[{{Damjanov} {et~al.}(2015){Damjanov}, {Geller}, {Zahid}, \&
  {Hwang}}]{damjanov15}
{Damjanov}, I., {Geller}, M.~J., {Zahid}, H.~J., \& {Hwang}, H.~S. 2015, \apj,
  806, 158

\bibitem[{{Damjanov} {et~al.}(2014){Damjanov}, {Hwang}, {Geller}, \&
  {Chilingarian}}]{damjanov14}
{Damjanov}, I., {Hwang}, H.~S., {Geller}, M.~J., \& {Chilingarian}, I. 2014,
  \apj, 793, 39

\bibitem[{{Damjanov} {et~al.}(2009){Damjanov}, {McCarthy}, {Abraham},
  {Glazebrook}, {Yan}, {Mentuch}, {Le Borgne}, {Savaglio}, {Crampton},
  {Murowinski}, {Juneau}, {Carlberg}, {J{\o}rgensen}, {Roth}, {Chen}, \&
  {Marzke}}]{damjanov09}
{Damjanov}, I., {McCarthy}, P.~J., {Abraham}, R.~G., {et~al.} 2009, \apj, 695,
  101

\bibitem[{{Davidzon} {et~al.}(2016){Davidzon}, {Cucciati}, {Bolzonella}, {De
  Lucia}, {Zamorani}, {Arnouts}, {Moutard}, {Ilbert}, {Garilli}, {Scodeggio},
  {Guzzo}, {Abbas}, {Adami}, {Bel}, {Bottini}, {Branchini}, {Cappi}, {Coupon},
  {de la Torre}, {Di Porto}, {Fritz}, {Franzetti}, {Fumana}, {Granett},
  {Guennou}, {Iovino}, {Krywult}, {Le Brun}, {Le F{\`e}vre}, {Maccagni},
  {Ma{\l}ek}, {Marulli}, {McCracken}, {Mellier}, {Moscardini}, {Polletta},
  {Pollo}, {Tasca}, {Tojeiro}, {Vergani}, \& {Zanichelli}}]{davidzon16}
{Davidzon}, I., {Cucciati}, O., {Bolzonella}, M., {et~al.} 2016, \aap, 586, A23

\bibitem[{{Donevski} {et~al.}(2020){Donevski}, {Lapi}, {Ma{\l}ek}, {Liu},
  {G{\'o}mez-Guijarro}, {Dav{\'e}}, {Kraljic}, {Pantoni}, {Man}, {Fujimoto},
  {Feltre}, {Pearson}, {Li}, \& {Narayanan}}]{donevski20}
{Donevski}, D., {Lapi}, A., {Ma{\l}ek}, K., {et~al.} 2020, \aap, 644, A144

\bibitem[{{Elbaz} {et~al.}(2018){Elbaz}, {Leiton}, {Nagar}, {Okumura},
  {Franco}, {Schreiber}, {Pannella}, {Wang}, {Dickinson}, {D{\'\i}az-Santos},
  {Ciesla}, {Daddi}, {Bournaud}, {Magdis}, {Zhou}, \& {Rujopakarn}}]{elbaz18}
{Elbaz}, D., {Leiton}, R., {Nagar}, N., {et~al.} 2018, \aap, 616, A110

\bibitem[{{Ferr{\'e}-Mateu} {et~al.}(2017){Ferr{\'e}-Mateu}, {Trujillo},
  {Mart{\'\i}n-Navarro}, {Vazdekis}, {Mezcua}, {Balcells}, \&
  {Dom{\'\i}nguez}}]{ferre-mateu17}
{Ferr{\'e}-Mateu}, A., {Trujillo}, I., {Mart{\'\i}n-Navarro}, I., {et~al.}
  2017, \mnras, 467, 1929

\bibitem[{{Flores-Freitas} {et~al.}(2021){Flores-Freitas}, {Chies-Santos},
  {Furlanetto}, {De Rossi}, {Ferreira}, {Zenocratti}, \&
  {Alamo-Mart{\'\i}nez}}]{flores-freitas21}
{Flores-Freitas}, R., {Chies-Santos}, A.~L., {Furlanetto}, C., {et~al.} 2021,
  arXiv e-prints, arXiv:2112.12846

\bibitem[{{Fritz} {et~al.}(2014){Fritz}, {Scodeggio}, {Ilbert}, {Bolzonella},
  {Davidzon}, {Coupon}, {Garilli}, {Guzzo}, {Zamorani}, {Abbas}, {Adami},
  {Arnouts}, {Bel}, {Bottini}, {Branchini}, {Cappi}, {Cucciati}, {De Lucia},
  {de la Torre}, {Franzetti}, {Fumana}, {Granett}, {Iovino}, {Krywult}, {Le
  Brun}, {Le F{\`e}vre}, {Maccagni}, {Ma{\l}ek}, {Marulli}, {McCracken},
  {Paioro}, {Polletta}, {Pollo}, {Schlagenhaufer}, {Tasca}, {Tojeiro},
  {Vergani}, {Zanichelli}, {Burden}, {Di Porto}, {Marchetti}, {Marinoni},
  {Mellier}, {Moscardini}, {Nichol}, {Peacock}, {Percival}, {Phleps}, \&
  {Wolk}}]{fritz14}
{Fritz}, A., {Scodeggio}, M., {Ilbert}, O., {et~al.} 2014, \aap, 563, A92

\bibitem[{{Furlong} {et~al.}(2017){Furlong}, {Bower}, {Crain}, {Schaye},
  {Theuns}, {Trayford}, {Qu}, {Schaller}, {Berthet}, \& {Helly}}]{furlong17}
{Furlong}, M., {Bower}, R.~G., {Crain}, R.~A., {et~al.} 2017, \mnras, 465, 722

\bibitem[{{Gargiulo} {et~al.}(2017){Gargiulo}, {Bolzonella}, {Scodeggio},
  {Krywult}, {De Lucia}, {Guzzo}, {Garilli}, {Granett}, {de la Torre}, {Abbas},
  {Adami}, {Arnouts}, {Bottini}, {Cappi}, {Cucciati}, {Davidzon}, {Franzetti},
  {Fritz}, {Haines}, {Hawken}, {Iovino}, {Le Brun}, {Le F{\`e}vre}, {Maccagni},
  {Ma{\l}ek}, {Marulli}, {Moutard}, {Polletta}, {Pollo}, {Tasca}, {Tojeiro},
  {Vergani}, {Zanichelli}, {Zamorani}, {Bel}, {Branchini}, {Coupon}, {Ilbert},
  {Moscardini}, \& {Peacock}}]{gargiulo17}
{Gargiulo}, A., {Bolzonella}, M., {Scodeggio}, M., {et~al.} 2017, \aap, 606,
  A113

\bibitem[{{Garilli} {et~al.}(2014){Garilli}, {Guzzo}, {Scodeggio},
  {Bolzonella}, {Abbas}, {Adami}, {Arnouts}, {Bel}, {Bottini}, {Branchini},
  {Cappi}, {Coupon}, {Cucciati}, {Davidzon}, {De Lucia}, {de la Torre},
  {Franzetti}, {Fritz}, {Fumana}, {Granett}, {Ilbert}, {Iovino}, {Krywult}, {Le
  Brun}, {Le F{\`e}vre}, {Maccagni}, {Ma{\l}ek}, {Marulli}, {McCracken},
  {Paioro}, {Polletta}, {Pollo}, {Schlagenhaufer}, {Tasca}, {Tojeiro},
  {Vergani}, {Zamorani}, {Zanichelli}, {Burden}, {Di Porto}, {Marchetti},
  {Marinoni}, {Mellier}, {Moscardini}, {Nichol}, {Peacock}, {Percival},
  {Phleps}, \& {Wolk}}]{garilli14}
{Garilli}, B., {Guzzo}, L., {Scodeggio}, M., {et~al.} 2014, \aap, 562, A23

\bibitem[{{Goranova, Y.} {et~al.}(2009){Goranova, Y.}, P., F., H., Y., M., M.,
  G., J.-C., \& H.}]{goranowa09}
{Goranova, Y.}, P., H., F., M., {et~al.} 2009, The CFHTLS T0006 Release, Tech.
  rep., Terapix/Institut d’Astrophysique de Paris

\bibitem[{{Graham} \& {Driver}(2005)}]{graham2005}
{Graham}, A.~W. \& {Driver}, S.~P. 2005, \pasa, 22, 118

\bibitem[{{Guzzo} {et~al.}(2014){Guzzo}, {Scodeggio}, {Garilli}, {Granett},
  {Fritz}, {Abbas}, {Adami}, {Arnouts}, {Bel}, {Bolzonella}, {Bottini},
  {Branchini}, {Cappi}, {Coupon}, {Cucciati}, {Davidzon}, {De Lucia}, {de la
  Torre}, {Franzetti}, {Fumana}, {Hudelot}, {Ilbert}, {Iovino}, {Krywult}, {Le
  Brun}, {Le F{\`e}vre}, {Maccagni}, {Ma{\l}ek}, {Marulli}, {McCracken},
  {Paioro}, {Peacock}, {Polletta}, {Pollo}, {Schlagenhaufer}, {Tasca},
  {Tojeiro}, {Vergani}, {Zamorani}, {Zanichelli}, {Burden}, {Di Porto},
  {Marchetti}, {Marinoni}, {Mellier}, {Moscardini}, {Nichol}, {Percival},
  {Phleps}, \& {Wolk}}]{guzzo14}
{Guzzo}, L., {Scodeggio}, M., {Garilli}, B., {et~al.} 2014, \aap, 566, A108

\bibitem[{{Haines} {et~al.}(2017){Haines}, {Iovino}, {Krywult}, {Guzzo},
  {Davidzon}, {Bolzonella}, {Garilli}, {Scodeggio}, {Granett}, {de la Torre},
  {De Lucia}, {Abbas}, {Adami}, {Arnouts}, {Bottini}, {Cappi}, {Cucciati},
  {Franzetti}, {Fritz}, {Gargiulo}, {Le Brun}, {Le F{\`e}vre}, {Maccagni},
  {Ma{\l}ek}, {Marulli}, {Moutard}, {Polletta}, {Pollo}, {Tasca}, {Tojeiro},
  {Vergani}, {Zanichelli}, {Zamorani}, {Bel}, {Branchini}, {Coupon}, {Ilbert},
  {Moscardini}, {Peacock}, \& {Siudek}}]{haines17}
{Haines}, C.~P., {Iovino}, A., {Krywult}, J., {et~al.} 2017, \aap, 605, A4

\bibitem[{{Hubble}(1926)}]{hubble26}
{Hubble}, E.~P. 1926, \apj, 64, 321

\bibitem[{{Ilbert} {et~al.}(2006){Ilbert}, {Arnouts}, {McCracken},
  {Bolzonella}, {Bertin}, {Le F{\`e}vre}, {Mellier}, {Zamorani}, {Pell{\`o}},
  {Iovino}, {Tresse}, {Le Brun}, {Bottini}, {Garilli}, {Maccagni}, {Picat},
  {Scaramella}, {Scodeggio}, {Vettolani}, {Zanichelli}, {Adami}, {Bardelli},
  {Cappi}, {Charlot}, {Ciliegi}, {Contini}, {Cucciati}, {Foucaud}, {Franzetti},
  {Gavignaud}, {Guzzo}, {Marano}, {Marinoni}, {Mazure}, {Meneux}, {Merighi},
  {Paltani}, {Pollo}, {Pozzetti}, {Radovich}, {Zucca}, {Bondi}, {Bongiorno},
  {Busarello}, {de La Torre}, {Gregorini}, {Lamareille}, {Mathez}, {Merluzzi},
  {Ripepi}, {Rizzo}, \& {Vergani}}]{ilbert06}
{Ilbert}, O., {Arnouts}, S., {McCracken}, H.~J., {et~al.} 2006, \aap, 457, 841

\bibitem[{{Ilbert} {et~al.}(2013){Ilbert}, {McCracken}, {Le F{\`e}vre},
  {Capak}, {Dunlop}, {Karim}, {Renzini}, {Caputi}, {Boissier}, {Arnouts},
  {Aussel}, {Comparat}, {Guo}, {Hudelot}, {Kartaltepe}, {Kneib}, {Krogager},
  {Le Floc'h}, {Lilly}, {Mellier}, {Milvang-Jensen}, {Moutard}, {Onodera},
  {Richard}, {Salvato}, {Sanders}, {Scoville}, {Silverman}, {Taniguchi},
  {Tasca}, {Thomas}, {Toft}, {Tresse}, {Vergani}, {Wolk}, \& {Zirm}}]{ilbert13}
{Ilbert}, O., {McCracken}, H.~J., {Le F{\`e}vre}, O., {et~al.} 2013, \aap, 556,
  A55

\bibitem[{{Jarvis} {et~al.}(2013){Jarvis}, {Bonfield}, {Bruce}, {Geach},
  {McAlpine}, {McLure}, {Gonz{\'a}lez-Solares}, {Irwin}, {Lewis}, {Yoldas},
  {Andreon}, {Cross}, {Emerson}, {Dalton}, {Dunlop}, {Hodgkin}, {Le},
  {Karouzos}, {Meisenheimer}, {Oliver}, {Rawlings}, {Simpson}, {Smail},
  {Smith}, {Sullivan}, {Sutherland}, {White}, \& {Zwart}}]{jarvis13}
{Jarvis}, M.~J., {Bonfield}, D.~G., {Bruce}, V.~A., {et~al.} 2013, \mnras, 428,
  1281

\bibitem[{{Kauffmann} {et~al.}(2003{\natexlab{a}}){Kauffmann}, {Heckman},
  {White}, {Charlot}, {Tremonti}, {Brinchmann}, {Bruzual}, {Peng}, {Seibert},
  {Bernardi}, {Blanton}, {Brinkmann}, {Castander}, {Cs{\'a}bai}, {Fukugita},
  {Ivezic}, {Munn}, {Nichol}, {Padmanabhan}, {Thakar}, {Weinberg}, \&
  {York}}]{kauffmann03}
{Kauffmann}, G., {Heckman}, T.~M., {White}, S. D.~M., {et~al.}
  2003{\natexlab{a}}, \mnras, 341, 33

\bibitem[{{Kauffmann} {et~al.}(2003{\natexlab{b}}){Kauffmann}, {Heckman},
  {White}, {Charlot}, {Tremonti}, {Brinchmann}, {Bruzual}, {Peng}, {Seibert},
  {Bernardi}, {Blanton}, {Brinkmann}, {Castander}, {Cs{\'a}bai}, {Fukugita},
  {Ivezic}, {Munn}, {Nichol}, {Padmanabhan}, {Thakar}, {Weinberg}, \&
  {York}}]{Kauffmann2003b}
{Kauffmann}, G., {Heckman}, T.~M., {White}, S. D.~M., {et~al.}
  2003{\natexlab{b}}, \mnras, 341, 33

\bibitem[{{Kennicutt}(1992)}]{kennicutt92}
{Kennicutt}, Robert~C., J. 1992, \apj, 388, 310

\bibitem[{{Kennicutt}(1998)}]{kennicut98}
{Kennicutt}, Robert~C., J. 1998, \araa, 36, 189

\bibitem[{{Krywult} {et~al.}(2017){Krywult}, {Tasca}, {Pollo}, {Vergani},
  {Bolzonella}, {Davidzon}, {Iovino}, {Gargiulo}, {Haines}, {Scodeggio},
  {Guzzo}, {Zamorani}, {Garilli}, {Granett}, {de la Torre}, {Abbas}, {Adami},
  {Bottini}, {Cappi}, {Cucciati}, {Franzetti}, {Fritz}, {Le Brun}, {Le
  F{\`e}vre}, {Maccagni}, {Ma{\l}ek}, {Marulli}, {Polletta}, {Tojeiro},
  {Zanichelli}, {Arnouts}, {Bel}, {Branchini}, {Coupon}, {De Lucia}, {Ilbert},
  {McCracken}, {Moscardini}, \& {Takeuchi}}]{krywult17}
{Krywult}, J., {Tasca}, L.~A.~M., {Pollo}, A., {et~al.} 2017, \aap, 598, A120

\bibitem[{{Le F{\`e}vre} {et~al.}(2003){Le F{\`e}vre}, {Saisse}, {Mancini},
  {Brau-Nogue}, {Caputi}, {Castinel}, {D'Odorico}, {Garilli}, {Kissler-Patig},
  {Lucuix}, {Mancini}, {Pauget}, {Sciarretta}, {Scodeggio}, {Tresse}, \&
  {Vettolani}}]{lefevre03}
{Le F{\`e}vre}, O., {Saisse}, M., {Mancini}, D., {et~al.} 2003, in Society of
  Photo-Optical Instrumentation Engineers (SPIE) Conference Series, Vol. 4841,
  Instrument Design and Performance for Optical/Infrared Ground-based
  Telescopes, ed. M.~{Iye} \& A.~F.~M. {Moorwood}, 1670--1681

\bibitem[{{Ma{\l}ek} {et~al.}(2018){Ma{\l}ek}, {Buat}, {Roehlly}, {Burgarella},
  {Hurley}, {Shirley}, {Duncan}, {Efstathiou}, {Papadopoulos}, {Vaccari},
  {Farrah}, {Marchetti}, \& {Oliver}}]{malek18}
{Ma{\l}ek}, K., {Buat}, V., {Roehlly}, Y., {et~al.} 2018, \aap, 620, A50

\bibitem[{{Moutard} {et~al.}(2016{\natexlab{a}}){Moutard}, {Arnouts}, {Ilbert},
  {Coupon}, {Davidzon}, {Guzzo}, {Hudelot}, {McCracken}, {Van Waerbeke},
  {Morrison}, {Le F{\`e}vre}, {Comte}, {Bolzonella}, {Fritz}, {Garilli}, \&
  {Scodeggio}}]{moutard16_II}
{Moutard}, T., {Arnouts}, S., {Ilbert}, O., {et~al.} 2016{\natexlab{a}}, \aap,
  590, A103

\bibitem[{{Moutard} {et~al.}(2016{\natexlab{b}}){Moutard}, {Arnouts}, {Ilbert},
  {Coupon}, {Hudelot}, {Vibert}, {Comte}, {Conseil}, {Davidzon}, {Guzzo},
  {Llebaria}, {Martin}, {McCracken}, {Milliard}, {Morrison}, {Schiminovich},
  {Treyer}, \& {Van Werbaeke}}]{moutard16_I}
{Moutard}, T., {Arnouts}, S., {Ilbert}, O., {et~al.} 2016{\natexlab{b}}, \aap,
  590, A102

\bibitem[{{Moutard} {et~al.}(2018){Moutard}, {Sawicki}, {Arnouts}, {Golob},
  {Malavasi}, {Adami}, {Coupon}, \& {Ilbert}}]{moutard18}
{Moutard}, T., {Sawicki}, M., {Arnouts}, S., {et~al.} 2018, \mnras, 479, 2147

\bibitem[{{Naab} {et~al.}(2009){Naab}, {Johansson}, \& {Ostriker}}]{naab09}
{Naab}, T., {Johansson}, P.~H., \& {Ostriker}, J.~P. 2009, \apjl, 699, L178

\bibitem[{{Oser} {et~al.}(2010){Oser}, {Ostriker}, {Naab}, {Johansson}, \&
  {Burkert}}]{oser10}
{Oser}, L., {Ostriker}, J.~P., {Naab}, T., {Johansson}, P.~H., \& {Burkert}, A.
  2010, \apj, 725, 2312

\bibitem[{{Peng} {et~al.}(2002){Peng}, {Ho}, {Impey}, \& {Rix}}]{peng02}
{Peng}, C.~Y., {Ho}, L.~C., {Impey}, C.~D., \& {Rix}, H.-W. 2002, \aj, 124, 266

\bibitem[{{Pistis} {et~al.}(2022){Pistis}, {Pollo}, {Scodeggio}, {Figueira},
  {Durkalec}, {Ma{\l}ek}, {Iovino}, {Vergani}, \& {Salim}}]{pastis22}
{Pistis}, F., {Pollo}, A., {Scodeggio}, M., {et~al.} 2022, arXiv e-prints,
  arXiv:2206.02458

\bibitem[{{Poggianti} {et~al.}(2013){Poggianti}, {Moretti}, {Calvi},
  {D'Onofrio}, {Valentinuzzi}, {Fritz}, \& {Renzini}}]{poggianti13}
{Poggianti}, B.~M., {Moretti}, A., {Calvi}, R., {et~al.} 2013, \apj, 777, 125

\bibitem[{{Pozzetti} {et~al.}(2010){Pozzetti}, {Bolzonella}, {Zucca},
  {Zamorani}, {Lilly}, {Renzini}, {Moresco}, {Mignoli}, {Cassata}, {Tasca},
  {Lamareille}, {Maier}, {Meneux}, {Halliday}, {Oesch}, {Vergani}, {Caputi},
  {Kova{\v{c}}}, {Cimatti}, {Cucciati}, {Iovino}, {Peng}, {Carollo}, {Contini},
  {Kneib}, {Le F{\'e}vre}, {Mainieri}, {Scodeggio}, {Bardelli}, {Bongiorno},
  {Coppa}, {de la Torre}, {de Ravel}, {Franzetti}, {Garilli}, {Kampczyk},
  {Knobel}, {Le Borgne}, {Le Brun}, {Pell{\`o}}, {Perez Montero},
  {Ricciardelli}, {Silverman}, {Tanaka}, {Tresse}, {Abbas}, {Bottini}, {Cappi},
  {Guzzo}, {Koekemoer}, {Leauthaud}, {Maccagni}, {Marinoni}, {McCracken},
  {Memeo}, {Porciani}, {Scaramella}, {Scarlata}, \& {Scoville}}]{pozzetti10}
{Pozzetti}, L., {Bolzonella}, M., {Zucca}, E., {et~al.} 2010, \aap, 523, A13

\bibitem[{{Quilis} \& {Trujillo}(2013)}]{quilis13}
{Quilis}, V. \& {Trujillo}, I. 2013, \apjl, 773, L8

\bibitem[{{Ra{\l}owski} {et~al.}(2020){Ra{\l}owski}, {Ma{\l}ek}, \&
  {Pollo}}]{ralowski20}
{Ra{\l}owski}, M., {Ma{\l}ek}, K., \& {Pollo}, A. 2020, in XXXIX Polish
  Astronomical Society Meeting, ed. K.~{Ma{\l}ek}, M.~{Poli{\'n}ska},
  A.~{Majczyna}, G.~{Stachowski}, R.~{Poleski}, {\L}.~{Wyrzykowski}, \&
  A.~{R{\'o}{\.z}a{\'n}ska}, Vol.~10, 231--236

\bibitem[{{Salim} {et~al.}(2018){Salim}, {Boquien}, \& {Lee}}]{salim18}
{Salim}, S., {Boquien}, M., \& {Lee}, J.~C. 2018, \apj, 859, 11

\bibitem[{{Saulder} {et~al.}(2015){Saulder}, {van den Bosch}, \&
  {Mieske}}]{saulder15}
{Saulder}, C., {van den Bosch}, R. C.~E., \& {Mieske}, S. 2015, \aap, 578, A134

\bibitem[{{Schreiber} {et~al.}(2015){Schreiber}, {Pannella}, {Elbaz},
  {B{\'e}thermin}, {Inami}, {Dickinson}, {Magnelli}, {Wang}, {Aussel}, {Daddi},
  {Juneau}, {Shu}, {Sargent}, {Buat}, {Faber}, {Ferguson}, {Giavalisco},
  {Koekemoer}, {Magdis}, {Morrison}, {Papovich}, {Santini}, \&
  {Scott}}]{schreiber15}
{Schreiber}, C., {Pannella}, M., {Elbaz}, D., {et~al.} 2015, \aap, 575, A74

\bibitem[{{Scodeggio} {et~al.}(2018){Scodeggio}, {Guzzo}, {Garilli}, {Granett},
  {Bolzonella}, {de la Torre}, {Abbas}, {Adami}, {Arnouts}, {Bottini}, {Cappi},
  {Coupon}, {Cucciati}, {Davidzon}, {Franzetti}, {Fritz}, {Iovino}, {Krywult},
  {Le Brun}, {Le F{\`e}vre}, {Maccagni}, {Ma{\l}ek}, {Marchetti}, {Marulli},
  {Polletta}, {Pollo}, {Tasca}, {Tojeiro}, {Vergani}, {Zanichelli}, {Bel},
  {Branchini}, {De Lucia}, {Ilbert}, {McCracken}, {Moutard}, {Peacock},
  {Zamorani}, {Burden}, {Fumana}, {Jullo}, {Marinoni}, {Mellier}, {Moscardini},
  \& {Percival}}]{scodeggio18}
{Scodeggio}, M., {Guzzo}, L., {Garilli}, B., {et~al.} 2018, \aap, 609, A84

\bibitem[{{Scognamiglio} {et~al.}(2020){Scognamiglio}, {Tortora}, {Spavone},
  {Spiniello}, {Napolitano}, {D'Ago}, {La Barbera}, {Getman}, {Roy}, {Raj},
  {Radovich}, {Brescia}, {Cavuoti}, {Koopmans}, {Kuijken}, {Longo}, \&
  {Petrillo}}]{scognamiglio20}
{Scognamiglio}, D., {Tortora}, C., {Spavone}, M., {et~al.} 2020, \apj, 893, 4

\bibitem[{{S{\'e}rsic}(1963)}]{sersic63}
{S{\'e}rsic}, J.~L. 1963, Boletin de la Asociacion Argentina de Astronomia La
  Plata Argentina, 6, 41

\bibitem[{{Sersic}(1968)}]{sersic68}
{Sersic}, J.~L. 1968, {Atlas de Galaxias Australes}

\bibitem[{{Siudek} {et~al.}(2018{\natexlab{a}}){Siudek}, {Ma{\l}ek}, {Pollo},
  {Granett}, {Scodeggio}, {Moutard}, {Iovino}, {Guzzo}, {Garilli},
  {Bolzonella}, {de la Torre}, {Abbas}, {Adami}, {Bottini}, {Cappi},
  {Cucciati}, {Davidzon}, {Franzetti}, {Fritz}, {Krywult}, {Le Brun}, {Le
  F{\`e}vre}, {Maccagni}, {Marulli}, {Polletta}, {Tasca}, {Tojeiro}, {Vergani},
  {Zanichelli}, {Arnouts}, {Bel}, {Branchini}, {Coupon}, {De Lucia}, {Ilbert},
  {Moscardini}, {Zamorani}, \& {Takeuchi}}]{siudek18_arx}
{Siudek}, M., {Ma{\l}ek}, K., {Pollo}, A., {et~al.} 2018{\natexlab{a}}, arXiv
  e-prints, arXiv:1805.09905

\bibitem[{{Siudek} {et~al.}(2022){Siudek}, {Malek}, {Pollo}, {Iovino},
  {Haines}, {Bolzonella}, {Cucciati}, {Gargiulo}, {Granett}, {Krywult},
  {Moutard}, \& {Scodeggio}}]{siudek22}
{Siudek}, M., {Malek}, K., {Pollo}, A., {et~al.} 2022, arXiv e-prints,
  arXiv:2205.14736

\bibitem[{{Siudek} {et~al.}(2018{\natexlab{b}}){Siudek}, {Ma{\l}ek}, {Pollo},
  {Krakowski}, {Iovino}, {Scodeggio}, {Moutard}, {Zamorani}, {Guzzo},
  {Garilli}, {Granett}, {Bolzonella}, {de la Torre}, {Abbas}, {Adami},
  {Bottini}, {Cappi}, {Cucciati}, {Davidzon}, {Franzetti}, {Fritz}, {Krywult},
  {Le Brun}, {Le F{\`e}vre}, {Maccagni}, {Marulli}, {Polletta}, {Tasca},
  {Tojeiro}, {Vergani}, {Zanichelli}, {Arnouts}, {Bel}, {Branchini}, {Coupon},
  {De Lucia}, {Ilbert}, {Haines}, {Moscardini}, \& {Takeuchi}}]{siudek18}
{Siudek}, M., {Ma{\l}ek}, K., {Pollo}, A., {et~al.} 2018{\natexlab{b}}, \aap,
  617, A70

\bibitem[{{Siudek} {et~al.}(2017){Siudek}, {Ma{\l}ek}, {Scodeggio}, {Garilli},
  {Pollo}, {Haines}, {Fritz}, {Bolzonella}, {de la Torre}, {Granett}, {Guzzo},
  {Abbas}, {Adami}, {Bottini}, {Cappi}, {Cucciati}, {De Lucia}, {Davidzon},
  {Franzetti}, {Iovino}, {Krywult}, {Le Brun}, {Le F{\`e}vre}, {Maccagni},
  {Marchetti}, {Marulli}, {Polletta}, {Tasca}, {Tojeiro}, {Vergani},
  {Zanichelli}, {Arnouts}, {Bel}, {Branchini}, {Ilbert}, {Gargiulo},
  {Moscardini}, {Takeuchi}, \& {Zamorani}}]{Siudek2017}
{Siudek}, M., {Ma{\l}ek}, K., {Scodeggio}, M., {et~al.} 2017, \aap, 597, A107

\bibitem[{{Spiniello} {et~al.}(2021){Spiniello}, {Tortora}, {D'Ago}, {Coccato},
  {La Barbera}, {Ferr{\'e}-Mateu}, {Napolitano}, {Spavone}, {Scognamiglio},
  {Arnaboldi}, {Gallazzi}, {Hunt}, {Moehler}, {Radovich}, \&
  {Zibetti}}]{spiniello21}
{Spiniello}, C., {Tortora}, C., {D'Ago}, G., {et~al.} 2021, \aap, 646, A28

\bibitem[{{Taylor} {et~al.}(2010){Taylor}, {Franx}, {Glazebrook}, {Brinchmann},
  {van der Wel}, \& {van Dokkum}}]{taylor10}
{Taylor}, E.~N., {Franx}, M., {Glazebrook}, K., {et~al.} 2010, \apj, 720, 723

\bibitem[{{Tortora} {et~al.}(2016){Tortora}, {La Barbera}, {Napolitano}, {Roy},
  {Radovich}, {Cavuoti}, {Brescia}, {Longo}, {Getman}, {Capaccioli}, {Grado},
  {Kuijken}, {de Jong}, {McFarland}, \& {Puddu}}]{tortora16}
{Tortora}, C., {La Barbera}, F., {Napolitano}, N.~R., {et~al.} 2016, \mnras,
  457, 2845

\bibitem[{{Trujillo} {et~al.}(2009){Trujillo}, {Cenarro}, {de
  Lorenzo-C{\'a}ceres}, {Vazdekis}, {de la Rosa}, \& {Cava}}]{trujillo09}
{Trujillo}, I., {Cenarro}, A.~J., {de Lorenzo-C{\'a}ceres}, A., {et~al.} 2009,
  \apjl, 692, L118

\bibitem[{{Trujillo} {et~al.}(2007){Trujillo}, {Conselice}, {Bundy}, {Cooper},
  {Eisenhardt}, \& {Ellis}}]{trujillo07}
{Trujillo}, I., {Conselice}, C.~J., {Bundy}, K., {et~al.} 2007, \mnras, 382,
  109

\bibitem[{{Turner} {et~al.}(2021){Turner}, {Siudek}, {Salim}, {Baldry},
  {Pollo}, {Longmore}, {Malek}, {Collins}, {Lisboa}, {Krywult}, {Moutard},
  {Vergani}, \& {Fritz}}]{turner21}
{Turner}, S., {Siudek}, M., {Salim}, S., {et~al.} 2021, \mnras, 503, 3010

\bibitem[{{Valentinuzzi} {et~al.}(2010){Valentinuzzi}, {Fritz}, {Poggianti},
  {Cava}, {Bettoni}, {Fasano}, {D'Onofrio}, {Couch}, {Dressler}, {Moles},
  {Moretti}, {Omizzolo}, {Kj{\ae}rgaard}, {Vanzella}, \&
  {Varela}}]{valentinuzzi10}
{Valentinuzzi}, T., {Fritz}, J., {Poggianti}, B.~M., {et~al.} 2010, \apj, 712,
  226

\bibitem[{{van der Wel} {et~al.}(2014){van der Wel}, {Franx}, {van Dokkum},
  {Skelton}, {Momcheva}, {Whitaker}, {Brammer}, {Bell}, {Rix}, {Wuyts},
  {Ferguson}, {Holden}, {Barro}, {Koekemoer}, {Chang}, {McGrath},
  {H{\"a}ussler}, {Dekel}, {Behroozi}, {Fumagalli}, {Leja}, {Lundgren},
  {Maseda}, {Nelson}, {Wake}, {Patel}, {Labb{\'e}}, {Faber}, {Grogin}, \&
  {Kocevski}}]{vanderwel14}
{van der Wel}, A., {Franx}, M., {van Dokkum}, P.~G., {et~al.} 2014, \apj, 788,
  28

\bibitem[{{van Dokkum} {et~al.}(2015){van Dokkum}, {Nelson}, {Franx}, {Oesch},
  {Momcheva}, {Brammer}, {F{\"o}rster Schreiber}, {Skelton}, {Whitaker}, {van
  der Wel}, {Bezanson}, {Fumagalli}, {Illingworth}, {Kriek}, {Leja}, \&
  {Wuyts}}]{vandokkum15}
{van Dokkum}, P.~G., {Nelson}, E.~J., {Franx}, M., {et~al.} 2015, \apj, 813, 23

\bibitem[{{van Dokkum} {et~al.}(2010){van Dokkum}, {Whitaker}, {Brammer},
  {Franx}, {Kriek}, {Labb{\'e}}, {Marchesini}, {Quadri}, {Bezanson},
  {Illingworth}, {Muzzin}, {Rudnick}, {Tal}, \& {Wake}}]{vandokkum10}
{van Dokkum}, P.~G., {Whitaker}, K.~E., {Brammer}, G., {et~al.} 2010, \apj,
  709, 1018

\bibitem[{{Vietri} {et~al.}(2021){Vietri}, {Garilli}, {Polletta}, {Bisogni},
  {Cassar{\`a}}, {Franzetti}, {Fumana}, {Gargiulo}, {Maccagni}, {Mancini},
  {Scodeggio}, {Fritz}, {Malek}, {Manzoni}, {Pollo}, {Siudek}, {Vergani},
  {Zamorani}, \& {Zanichelli}}]{Vietri2021}
{Vietri}, G., {Garilli}, B., {Polletta}, M., {et~al.} 2021, arXiv e-prints,
  arXiv:2111.08730

\bibitem[{{Wellons} {et~al.}(2016){Wellons}, {Torrey}, {Ma}, {Rodriguez-Gomez},
  {Pillepich}, {Nelson}, {Genel}, {Vogelsberger}, \& {Hernquist}}]{wellons16}
{Wellons}, S., {Torrey}, P., {Ma}, C.-P., {et~al.} 2016, \mnras, 456, 1030

\bibitem[{{Wright} {et~al.}(2010){Wright}, {Eisenhardt}, {Mainzer}, {Ressler},
  {Cutri}, {Jarrett}, {Kirkpatrick}, {Padgett}, {McMillan}, {Skrutskie},
  {Stanford}, {Cohen}, {Walker}, {Mather}, {Leisawitz}, {Gautier}, {McLean},
  {Benford}, {Lonsdale}, {Blain}, {Mendez}, {Irace}, {Duval}, {Liu}, {Royer},
  {Heinrichsen}, {Howard}, {Shannon}, {Kendall}, {Walsh}, {Larsen}, {Cardon},
  {Schick}, {Schwalm}, {Abid}, {Fabinsky}, {Naes}, \& {Tsai}}]{wright10}
{Wright}, E.~L., {Eisenhardt}, P. R.~M., {Mainzer}, A.~K., {et~al.} 2010, \aj,
  140, 1868

\bibitem[{{York} {et~al.}(2000){York}, {Adelman}, {Anderson}, {Anderson},
  {Annis}, {Bahcall}, {Bakken}, {Barkhouser}, {Bastian}, {Berman}, {Boroski},
  {Bracker}, {Briegel}, {Briggs}, {Brinkmann}, {Brunner}, {Burles}, {Carey},
  {Carr}, {Castander}, {Chen}, {Colestock}, {Connolly}, {Crocker}, {Csabai},
  {Czarapata}, {Davis}, {Doi}, {Dombeck}, {Eisenstein}, {Ellman}, {Elms},
  {Evans}, {Fan}, {Federwitz}, {Fiscelli}, {Friedman}, {Frieman}, {Fukugita},
  {Gillespie}, {Gunn}, {Gurbani}, {de Haas}, {Haldeman}, {Harris}, {Hayes},
  {Heckman}, {Hennessy}, {Hindsley}, {Holm}, {Holmgren}, {Huang}, {Hull},
  {Husby}, {Ichikawa}, {Ichikawa}, {Ivezi{\'c}}, {Kent}, {Kim}, {Kinney},
  {Klaene}, {Kleinman}, {Kleinman}, {Knapp}, {Korienek}, {Kron}, {Kunszt},
  {Lamb}, {Lee}, {Leger}, {Limmongkol}, {Lindenmeyer}, {Long}, {Loomis},
  {Loveday}, {Lucinio}, {Lupton}, {MacKinnon}, {Mannery}, {Mantsch}, {Margon},
  {McGehee}, {McKay}, {Meiksin}, {Merelli}, {Monet}, {Munn}, {Narayanan},
  {Nash}, {Neilsen}, {Neswold}, {Newberg}, {Nichol}, {Nicinski}, {Nonino},
  {Okada}, {Okamura}, {Ostriker}, {Owen}, {Pauls}, {Peoples}, {Peterson},
  {Petravick}, {Pier}, {Pope}, {Pordes}, {Prosapio}, {Rechenmacher}, {Quinn},
  {Richards}, {Richmond}, {Rivetta}, {Rockosi}, {Ruthmansdorfer}, {Sandford},
  {Schlegel}, {Schneider}, {Sekiguchi}, {Sergey}, {Shimasaku}, {Siegmund},
  {Smee}, {Smith}, {Snedden}, {Stone}, {Stoughton}, {Strauss}, {Stubbs},
  {SubbaRao}, {Szalay}, {Szapudi}, {Szokoly}, {Thakar}, {Tremonti}, {Tucker},
  {Uomoto}, {Vanden Berk}, {Vogeley}, {Waddell}, {Wang}, {Watanabe},
  {Weinberg}, {Yanny}, {Yasuda}, \& {SDSS Collaboration}}]{york00}
{York}, D.~G., {Adelman}, J., {Anderson}, John~E., J., {et~al.} 2000, \aj, 120,
  1579

\bibitem[{{Zibetti} {et~al.}(2020){Zibetti}, {Gallazzi}, {Hirschmann},
  {Consolandi}, {Falc{\'o}n-Barroso}, {van de Ven}, \& {Lyubenova}}]{zibetti20}
{Zibetti}, S., {Gallazzi}, A.~R., {Hirschmann}, M., {et~al.} 2020, \mnras, 491,
  3562

\bibitem[{{Zolotov} {et~al.}(2015){Zolotov}, {Dekel}, {Mandelker}, {Tweed},
  {Inoue}, {DeGraf}, {Ceverino}, {Primack}, {Barro}, \& {Faber}}]{zolotov15}
{Zolotov}, A., {Dekel}, A., {Mandelker}, N., {et~al.} 2015, \mnras, 450, 2327

\end{thebibliography}

\appendix
\section{Attenuation}
\label{app:attenuation}
\begin{figure*}[ht]
\centering
\includegraphics[width = 1\textwidth]{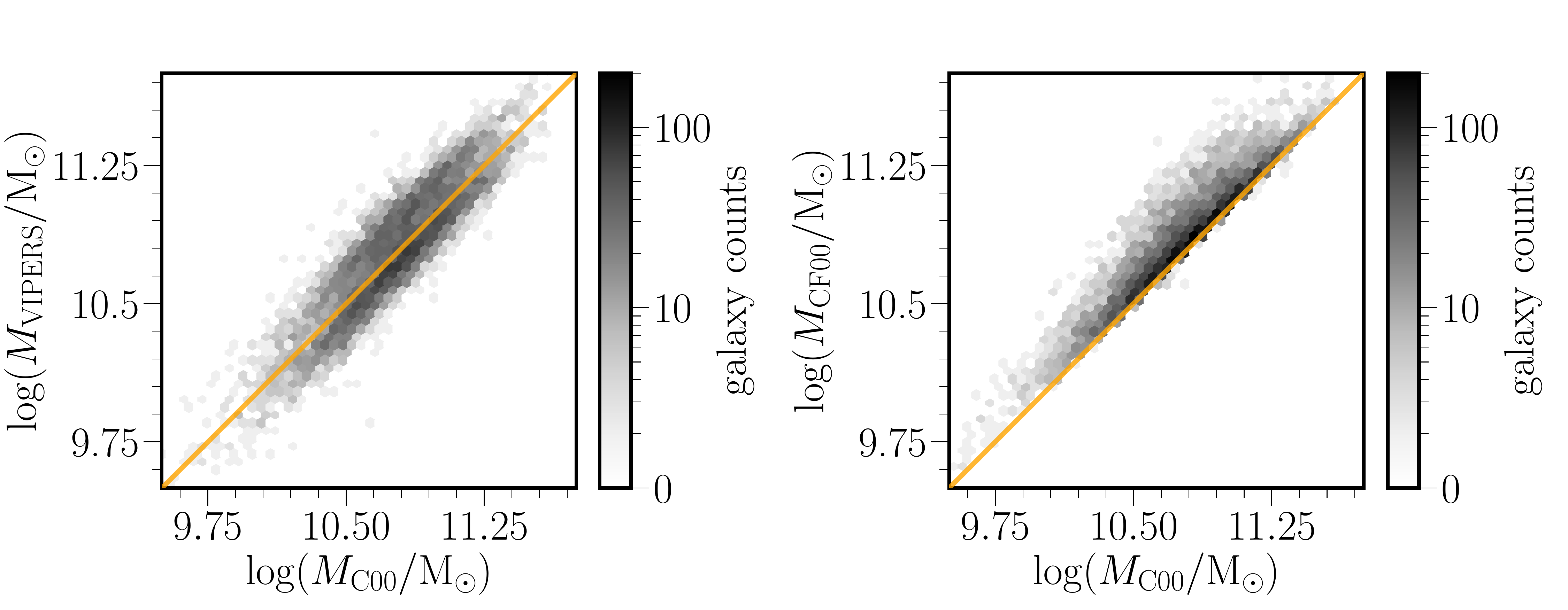}
\caption{The left panel shows comparison of stellar masses calculated via CIGALE with \cite{Calzetti2000} attenuation law, $M_{\rm{C00}}$,  with those calculated by the VIPERS team using Le Phare tool \citep{arnouts99, ilbert06}, $ M_{\rm VIPERS}$ for 36\,157 galaxies in the \textit{pure sample}. The right panel shows comparison of stellar masses calculated with CIGALE using \cite{Calzetti2000} and \cite{CF2000} $M_{\rm CF00}$ attenuation laws  for the same \textit{pure sample} of galaxies. On both plots one to one relation was marked with an orange line.}
\label{fig:mstar_comparison}
\end{figure*}

\begin{table*}[ht]
\centering
	\caption{Input parameters used in SED fitting with CIGALE.}
	\label{CF00_initial}
	\footnotesize
	\begin{tabular}{c c}
\hline
\hline
Parameter & Values\\
\hline
\multicolumn{2}{c}{Delayed star formation history}\\
\hline
e-folding time of the main stellar population model (Myr) & {100, 300, 500, 1500}\\
Age of the main stellar population& 300, 500, 800, 1000, 2500, \\
 in the galaxy (Myr) &3500, 4500, 6000, 7000\\
e-folding time of the late starburst population model (Myr) & 10000\\
Age of the late burst (Myr)& 10\\
Mass fraction of the late burst population & 0\\
Normalise the SFH to produce one solar mass & True\\
\hline
\multicolumn{2}{c}{Stellar population synthesis  \cite{bruzal03}}\\
\hline
Initial mass function & \cite{chabrier03}\\
Metalicity & 0.02\\
Age of the separation between the young and the old star populations (Myr) & 10\\
\hline
\multicolumn{2}{c}{Dust emission \citealp{dale14}}\\
\hline
AGN fraction & 0\\
$\alpha$ slope & 2.0\\
\hline
\multicolumn{2}{c}{Dust attenuation \cite{CF2000}}\\
\hline
V-band attenuation in the interstellar medium & 0.0, 0.05, 0.1, 0.2, 0.4, 0.75,\\
&1.0, 1.5, 1.75, 2.0, 2.2, 2.5, 3.0\\
$\mu$ & 0.8\\
Power law slope of the attenuation in the ISM & -0.7\\
Power law slope of the attenuation in the birth clouds & -0.7\\
\hline
\multicolumn{2}{c}{Dust attenuation \cite{Calzetti2000}}\\
\hline
E(B-V)l & 0.0,0.05, 0.1, 0.2, 0.3, 0.4, 0.5, \\
&0.6, 0.7, 0.8,0.9, 1.0 ,1.3, 1.5\\
E\_BV factor & 0.44\\
Central wavelength of the UV bump in nm & 217.5\\
Width (FWHM) of the UV bump in nm & 35.0\\
Amplitude of the UV bump & 0\\
Slope delta of the power law modifying the attenuation curve & 0\\
Extinction law to use for attenuating the emission lines flux & Milky Way\\
Ratio A\_V / E(B-V) & 3.1\\
\hline
\hline
\end{tabular}
\label{tab:cigale_input}
\end{table*}	

Figure~\ref{fig:mstar_comparison} shows that stellar masses estimated with the Calzetti attenuation law \citep{Calzetti2000} are much closer to the ones originally estimated using the Le Phare SED fitting tool \citep{arnouts99, ilbert06} by the VIPERS team. 
The mean scatter between our results obtained with the Calzetti attenuation law and the VIPERS stellar masses is less than 0.02 dex.
The median difference obtained with both attenuation laws is $\sim$0.1 dex. The Calzetti model is simpler and has fewer degrees of freedom 
than the double power law model of \cite{CF2000},
which assumes separate power laws for the birth cloud and one for the interstellar medium.
As we lack reliable IR data, this is an advantage in this case.
We therefore decided to use in our analysis results obtained with \cite{Calzetti2000} law.

\section{Examples of VIPERS' red nuggets}
\label{app:examples}

\begin{figure*}[ht]
\centering
\includegraphics[width = 0.9\textwidth]{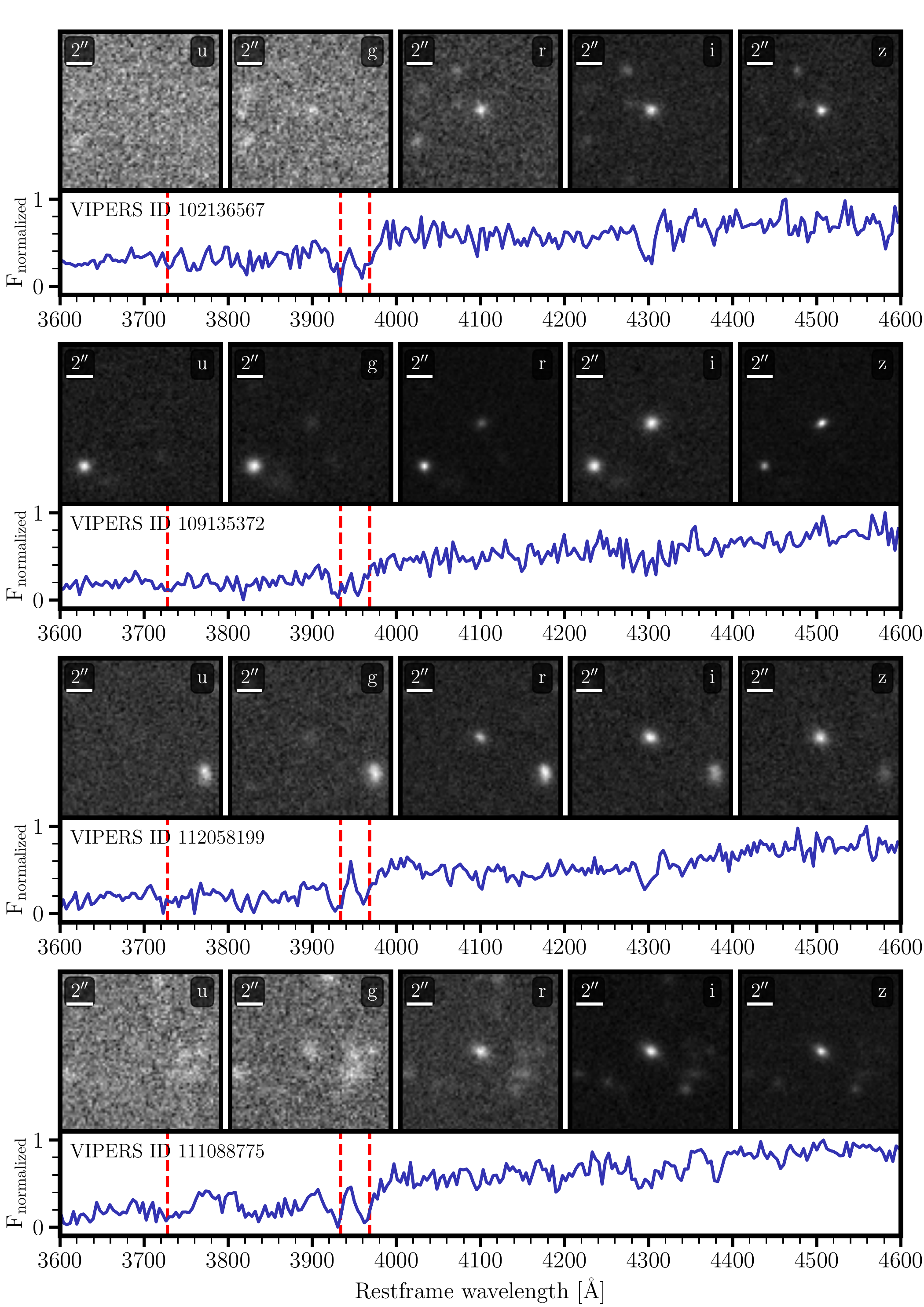}
\caption{Examples of red nuggets in our catalogue. For every galaxy we show images in the \textit{u}, \textit{g}, \textit{r}, \textit{i}, and \textit{z} bands from the CFHT survey, and normalized spectra with wavelength of oxygen [OII] emission line (3727.5 \AA), calcium CaII~K (3933.7 \AA) and H (3968.5 \AA) absorption lines marked. }
\label{fig:examples}
\end{figure*}

\onecolumn
\begingroup
\renewcommand{\arraystretch}{1.5} 
\section{Catalogue of red nuggets}\label{Sec:app_catalogue}

\begin{longtable}{c c c c c c c c }
\caption{Catalogue of VIPERS red nuggets. 
We present the main physical properties used in our analysis. The first column contains VIERS ID, the next two show the sky position, the third one shows redshift, Ra, Dec, $z$. 
The next two columns contain physical properties like circularised half-light radii, stellar mass respectively. 
The last two columns show the colours used for passiveness checking: NUV - r and r - K. Three VIPERS IDs, together with their R$_e$ values,  are underlined, as their relative errors of $R_e$ are higher than 45\%.}\\
\hline
VIPERS ID & Ra [deg] & Dec [deg] & $z$ & $R_{e}$ [kpc] & $\log(M_{star})$ [$\rm{M}_{\odot}$]& NUV - r& r - K\\
\hline
101131036 & \phantom{1}30.5258639 & -5.9354478 & 0.997 & 0.73 $\pm$ 0.11 & 10.99 $^{+0.03}_{-0.03}$ & 4.41 $\pm$ 0.09 & 0.99 $\pm$ 0.09 \\
102136567 & \phantom{1}31.5245661 & -5.8885126 & 0.927 & 1.19 $\pm$ 0.51 & 10.90 $^{+0.08}_{-0.09}$ & 5.39 $\pm$ 0.21 & 0.85 $\pm$ 0.09 \\
102164621 & \phantom{1}31.5018244 & -5.7478308 & 0.845 & 1.10 $\pm$ 0.13 & 10.91 $^{+0.07}_{-0.09}$ & 5.36 $\pm$ 0.16 & 0.83 $\pm$ 0.09 \\
103137838 & \phantom{1}32.7348014 & -5.9634442 & 0.750 & 1.31 $\pm$ 0.11 & 10.96 $^{+0.08}_{-0.10}$ & 5.26 $\pm$ 0.38 & 0.95 $\pm$ 0.10 \\
104169502 & \phantom{1}33.3173687 & -5.9113035 & 0.766 & 0.65 $\pm$ 0.04 & 10.95 $^{+0.05}_{-0.06}$ & 5.21 $\pm$ 0.25 & 0.88 $\pm$ 0.11 \\
104243009 & \phantom{1}33.5006282 & -5.6283503 & 0.742 & 1.37 $\pm$ 0.13 & 11.01 $^{+0.07}_{-0.08}$ & 6.03 $\pm$ 0.16 & 1.01 $\pm$ 0.10 \\
105149946 & \phantom{1}34.9739041 & -5.9108378 & 0.777 & 1.15 $\pm$ 0.07 & 10.98 $^{+0.07}_{-0.09}$ & 5.50 $\pm$ 0.18 & 0.84 $\pm$ 0.08 \\
106178573 & \phantom{1}35.4706881 & -5.7853115 & 0.776 & 1.25 $\pm$ 0.06 & 11.08 $^{+0.03}_{-0.03}$ & 5.19 $\pm$ 0.08 & 0.74 $\pm$ 0.08 \\
107109094 & \phantom{1}36.6137710 & -5.9471244 & 0.820 & 1.24 $\pm$ 0.08 & 10.91 $^{+0.04}_{-0.05}$ & 4.23 $\pm$ 0.11 & 0.91 $\pm$ 0.11 \\
107115227 & \phantom{1}36.5739924 & -5.9120644 & 0.978 & 1.42 $\pm$ 0.26 & 11.22 $^{+0.06}_{-0.07}$ & 5.33 $\pm$ 0.15 & 0.83 $\pm$ 0.08 \\
107156787 & \phantom{1}36.0857272 & -5.6749818 & 0.684 & 1.14 $\pm$ 0.09 & 10.93 $^{+0.05}_{-0.06}$ & 4.69 $\pm$ 0.25 & 0.89 $\pm$ 0.08 \\
109135372 & \phantom{1}37.9142200 & -5.8984831 & 0.833 & 0.84 $\pm$ 0.27 & 11.02 $^{+0.08}_{-0.09}$ & 5.56 $\pm$ 0.23 & 0.90 $\pm$ 0.09 \\
109179636 & \phantom{1}38.3647960 & -5.6751634 & 0.524 & 1.07 $\pm$ 0.21 & 11.28 $^{+0.08}_{-0.11}$ & 5.81 $\pm$ 0.17 & 1.01 $\pm$ 0.08 \\
110144150 & \phantom{1}30.3309573 & -4.9606058 & 0.764 & 1.22 $\pm$ 0.11 & 10.91 $^{+0.08}_{-0.10}$ & 5.66 $\pm$ 0.21 & 0.88 $\pm$ 0.08 \\
111088775 & \phantom{1}31.5152978 & -5.1940591 & 0.817 & 1.47 $\pm$ 0.14 & 10.98 $^{+0.08}_{-0.09}$ & 5.58 $\pm$ 0.20 & 0.88 $\pm$ 0.08 \\
111091579 & \phantom{1}31.4664203 & -5.1807426 & 0.666 & 1.27 $\pm$ 0.04 & 10.94 $^{+0.06}_{-0.07}$ & 5.26 $\pm$ 0.13 & 0.79 $\pm$ 0.08 \\
111126369 & \phantom{1}31.2330626 & -5.0174692 & 0.857 & 1.22 $\pm$ 0.11 & 10.98 $^{+0.08}_{-0.10}$ & 5.52 $\pm$ 0.20 & 0.87 $\pm$ 0.08 \\
112058199 & \phantom{1}32.8933163 & -5.3494238 & 0.920 & 1.46 $\pm$ 0.17 & 11.10 $^{+0.08}_{-0.10}$ & 5.45 $\pm$ 0.18 & 0.87 $\pm$ 0.08 \\
112117953 & \phantom{1}32.8196331 & -5.0724753 & 0.910 & 0.92 $\pm$ 0.16 & 10.97 $^{+0.02}_{-0.03}$ & 5.20 $\pm$ 0.09 & 0.73 $\pm$ 0.08 \\
112184594 & \phantom{1}32.7301990 & -4.7586513 & 0.752 & 1.44 $\pm$ 0.12 & 11.00 $^{+0.08}_{-0.10}$ & 5.78 $\pm$ 0.18 & 0.92 $\pm$ 0.08 \\
112195565 & \phantom{1}32.1520740 & -4.7060957 & 0.669 & 1.36 $\pm$ 0.08 & 11.07 $^{+0.08}_{-0.10}$ & 5.86 $\pm$ 0.28 & 1.02 $\pm$ 0.09 \\
114054377 & \phantom{1}34.8674224 & -5.3974010 & 0.907 & 0.52 $\pm$ 0.01 & 10.94 $^{+0.10}_{-0.13}$ & 5.12 $\pm$ 0.36 & 1.14 $\pm$ 0.09 \\
114188387 & \phantom{1}34.9725993 & -4.8450140 & 0.905 & 0.88 $\pm$ 0.11 & 11.00 $^{+0.05}_{-0.06}$ & 5.33 $\pm$ 0.18 & 0.84 $\pm$ 0.08 \\
115099121 & \phantom{1}35.3476297 & -5.1404602 & 0.902 & 0.93 $\pm$ 0.21 & 11.01 $^{+0.07}_{-0.09}$ & 5.69 $\pm$ 0.19 & 0.93 $\pm$ 0.08 \\
115108461 & \phantom{1}35.6193505 & -5.0970254 & 0.833 & 1.44 $\pm$ 0.19 & 11.04 $^{+0.07}_{-0.09}$ & 5.70 $\pm$ 0.19 & 0.92 $\pm$ 0.08 \\
117178615 & \phantom{1}37.5623537 & -4.7490847 & 0.938 & 1.24 $\pm$ 0.18 & 11.09 $^{+0.08}_{-0.10}$ & 5.47 $\pm$ 0.21 & 0.94 $\pm$ 0.15 \\
118035134 & \phantom{1}38.4941872 & -5.4508176 & 0.948 & 0.93 $\pm$ 0.18 & 10.91 $^{+0.09}_{-0.11}$ & 4.34 $\pm$ 0.11 & 0.91 $\pm$ 0.11 \\
\underline{118051171} & \phantom{1}38.6463163 & -5.3644373 & 0.725 & \underline{1.09 $\pm$ 0.68} & 10.94 $^{+0.09}_{-0.11}$ & 5.74 $\pm$ 0.20 & 0.91 $\pm$ 0.10 \\
118052178 & \phantom{1}38.5611866 & -5.3593346 & 0.937 & 1.07 $\pm$ 0.20 & 11.07 $^{+0.03}_{-0.04}$ & 5.22 $\pm$ 0.09 & 0.78 $\pm$ 0.08 \\
118054089 & \phantom{1}38.7315875 & -5.3489438 & 0.782 & 1.42 $\pm$ 0.15 & 10.99 $^{+0.11}_{-0.15}$ & 5.49 $\pm$ 0.20 & 0.89 $\pm$ 0.23 \\
118058527 & \phantom{1}38.6383557 & -5.3264219 & 0.742 & 1.38 $\pm$ 0.08 & 10.98 $^{+0.04}_{-0.04}$ & 5.18 $\pm$ 0.16 & 0.75 $\pm$ 0.08 \\
118058570 & \phantom{1}38.7665364 & -5.3267823 & 0.860 & 0.92 $\pm$ 0.14 & 10.91 $^{+0.09}_{-0.11}$ & 5.45 $\pm$ 0.17 & 0.86 $\pm$ 0.12 \\
119108256 & \phantom{1}30.9170954 & -4.1747378 & 0.816 & 1.11 $\pm$ 0.10 & 10.98 $^{+0.08}_{-0.10}$ & 5.42 $\pm$ 0.17 & 0.84 $\pm$ 0.11 \\
120083762 & \phantom{1}31.8664909 & -4.3497990 & 0.962 & 1.15 $\pm$ 0.15 & 10.93 $^{+0.08}_{-0.10}$ & 4.70 $\pm$ 0.23 & 0.99 $\pm$ 0.11 \\
121065375 & \phantom{1}33.0372818 & -4.3898337 & 0.967 & 0.79 $\pm$ 0.00 & 10.90 $^{+0.08}_{-0.10}$ & 4.49 $\pm$ 0.22 & 0.90 $\pm$ 0.09 \\
121066765 & \phantom{1}32.1371147 & -4.3834791 & 0.973 & 1.18 $\pm$ 0.18 & 11.02 $^{+0.04}_{-0.05}$ & 5.13 $\pm$ 0.20 & 0.79 $\pm$ 0.08 \\
121089481 & \phantom{1}32.1151895 & -4.2845186 & 0.780 & 1.39 $\pm$ 0.12 & 11.02 $^{+0.09}_{-0.11}$ & 5.30 $\pm$ 0.35 & 1.24 $\pm$ 0.12 \\
122030630 & \phantom{1}33.7481148 & -4.5429696 & 0.681 & 1.48 $\pm$ 0.07 & 10.95 $^{+0.08}_{-0.10}$ & 5.66 $\pm$ 0.26 & 0.87 $\pm$ 0.08 \\
122057921 & \phantom{1}33.5865511 & -4.4072717 & 0.699 & 1.25 $\pm$ 0.16 & 10.95 $^{+0.08}_{-0.10}$ & 5.77 $\pm$ 0.27 & 0.93 $\pm$ 0.10 \\
123089330 & \phantom{1}34.4349917 & -4.2838982 & 0.730 & 1.45 $\pm$ 0.01 & 10.93 $^{+0.07}_{-0.08}$ & 5.30 $\pm$ 0.37 & 0.88 $\pm$ 0.08 \\
124039182 & \phantom{1}35.4891458 & -4.5125297 & 0.834 & 1.16 $\pm$ 0.08 & 11.10 $^{+0.10}_{-0.13}$ & 4.30 $\pm$ 0.11 & 0.88 $\pm$ 0.08 \\
124039429 & \phantom{1}35.1412790 & -4.5114185 & 0.847 & 1.49 $\pm$ 0.11 & 10.91 $^{+0.03}_{-0.03}$ & 4.67 $\pm$ 0.13 & 1.15 $\pm$ 0.08 \\
124070857 & \phantom{1}35.6400158 & -4.3651440 & 0.894 & 1.47 $\pm$ 0.19 & 10.99 $^{+0.06}_{-0.07}$ & 5.15 $\pm$ 0.29 & 0.88 $\pm$ 0.08 \\
127044375 & \phantom{1}38.3579640 & -4.4846978 & 0.971 & 1.43 $\pm$ 0.33 & 10.99 $^{+0.06}_{-0.07}$ & 4.97 $\pm$ 0.32 & 0.80 $\pm$ 0.09 \\
\underline{401010173} & 330.7162868 & \phantom{-}0.8647405 & 0.758 & \underline{1.09 $\pm$ 1.08} & 10.95 $^{+0.05}_{-0.06}$ & 3.86 $\pm$ 0.09 & 0.76 $\pm$ 0.14 \\
401031898 & 330.7988766 & \phantom{-}0.9561888 & 0.697 & 1.04 $\pm$ 0.11 & 10.96 $^{+0.08}_{-0.09}$ & 6.19 $\pm$ 0.32 & 1.19 $\pm$ 0.14 \\
401174682 & 330.5012232 & \phantom{-}1.5673446 & 0.670 & 1.09 $\pm$ 0.08 & 10.93 $^{+0.08}_{-0.10}$ & 5.74 $\pm$ 0.19 & 0.89 $\pm$ 0.08 \\
401213761 & 330.5577447 & \phantom{-}1.7365884 & 0.845 & 0.83 $\pm$ 0.06 & 10.95 $^{+0.03}_{-0.04}$ & 5.15 $\pm$ 0.20 & 0.74 $\pm$ 0.08 \\
402034655 & 331.5967332 & \phantom{-}0.9564622 & 0.985 & 1.50 $\pm$ 0.09 & 11.16 $^{+0.04}_{-0.04}$ & 4.13 $\pm$ 0.08 & 0.79 $\pm$ 0.10 \\
402071577 & 331.7504873 & \phantom{-}1.1141104 & 0.557 & 1.22 $\pm$ 0.04 & 10.98 $^{+0.05}_{-0.06}$ & 5.27 $\pm$ 0.13 & 0.80 $\pm$ 0.08 \\
402075551 & 331.1296719 & \phantom{-}1.1303454 & 0.876 & 1.05 $\pm$ 0.10 & 10.91 $^{+0.08}_{-0.09}$ & 4.02 $\pm$ 0.11 & 0.78 $\pm$ 0.14 \\
402165870 & 331.2811478 & \phantom{-}1.4969730 & 0.990 & 1.22 $\pm$ 0.00 & 11.02 $^{+0.07}_{-0.08}$ & 4.78 $\pm$ 0.25 & 1.00 $\pm$ 0.12 \\
402173836 & 331.3283545 & \phantom{-}1.5284091 & 0.770 & 1.11 $\pm$ 0.06 & 11.00 $^{+0.09}_{-0.12}$ & 4.47 $\pm$ 0.21 & 0.89 $\pm$ 0.11 \\
403027664 & 332.5970905 & \phantom{-}0.9475146 & 0.682 & 1.32 $\pm$ 0.08 & 10.92 $^{+0.08}_{-0.11}$ & 5.73 $\pm$ 0.29 & 0.92 $\pm$ 0.09 \\
403033676 & 332.7095288 & \phantom{-}0.9741027 & 0.744 & 1.46 $\pm$ 0.08 & 11.09 $^{+0.10}_{-0.12}$ & 5.45 $\pm$ 0.44 & 1.16 $\pm$ 0.13 \\
403037148 & 332.3363850 & \phantom{-}0.9906283 & 0.894 & 1.01 $\pm$ 0.10 & 10.92 $^{+0.04}_{-0.05}$ & 5.17 $\pm$ 0.17 & 0.76 $\pm$ 0.08 \\
403054648 & 332.3993110 & \phantom{-}1.0743524 & 0.893 & 1.50 $\pm$ 0.16 & 10.97 $^{+0.08}_{-0.10}$ & 5.59 $\pm$ 0.28 & 0.92 $\pm$ 0.12 \\
403058809 & 332.3622401 & \phantom{-}1.0945229 & 0.888 & 1.41 $\pm$ 0.15 & 10.95 $^{+0.08}_{-0.10}$ & 5.61 $\pm$ 0.25 & 0.92 $\pm$ 0.12 \\
403081598 & 332.3105808 & \phantom{-}1.1951127 & 0.903 & 0.86 $\pm$ 0.10 & 10.92 $^{+0.09}_{-0.12}$ & 4.41 $\pm$ 0.18 & 0.89 $\pm$ 0.10 \\
403086972 & 332.6793812 & \phantom{-}1.2183092 & 0.716 & 0.65 $\pm$ 0.03 & 10.99 $^{+0.09}_{-0.12}$ & 4.41 $\pm$ 0.24 & 0.85 $\pm$ 0.09 \\
403099223 & 332.4299073 & \phantom{-}1.2729087 & 0.747 & 1.47 $\pm$ 0.10 & 10.95 $^{+0.09}_{-0.11}$ & 4.56 $\pm$ 0.34 & 0.79 $\pm$ 0.08 \\
403101616 & 332.5842966 & \phantom{-}1.2837705 & 0.895 & 1.34 $\pm$ 0.01 & 10.96 $^{+0.08}_{-0.10}$ & 5.37 $\pm$ 0.23 & 0.86 $\pm$ 0.12 \\
404066979 & 333.1519692 & \phantom{-}1.1316766 & 0.795 & 1.26 $\pm$ 0.25 & 10.95 $^{+0.07}_{-0.09}$ & 5.97 $\pm$ 0.27 & 1.09 $\pm$ 0.15 \\
404119465 & 333.2418082 & \phantom{-}1.3699357 & 0.931 & 1.40 $\pm$ 0.16 & 11.01 $^{+0.07}_{-0.09}$ & 5.76 $\pm$ 0.17 & 0.93 $\pm$ 0.10 \\
404139209 & 333.3773493 & \phantom{-}1.4592410 & 0.647 & 1.48 $\pm$ 0.09 & 10.96 $^{+0.07}_{-0.09}$ & 5.86 $\pm$ 0.18 & 0.92 $\pm$ 0.08 \\
405113585 & 333.8988503 & \phantom{-}1.3478597 & 0.927 & 0.95 $\pm$ 0.08 & 10.97 $^{+0.05}_{-0.05}$ & 5.09 $\pm$ 0.27 & 0.78 $\pm$ 0.11 \\
\underline{405183305} & 334.1749485 & \phantom{-}1.6714599 & 0.573 & \underline{1.30 $\pm$ 0.79} & 11.14 $^{+0.05}_{-0.06}$ & 4.50 $\pm$ 0.17 & 0.87 $\pm$ 0.08 \\
408062025 & 331.3608763 & \phantom{-}2.0556732 & 0.716 & 1.10 $\pm$ 0.10 & 10.96 $^{+0.09}_{-0.12}$ & 5.67 $\pm$ 0.38 & 0.97 $\pm$ 0.10 \\
408062039 & 331.1839421 & \phantom{-}2.0560387 & 0.618 & 1.48 $\pm$ 0.04 & 10.91 $^{+0.07}_{-0.08}$ & 5.31 $\pm$ 0.19 & 0.83 $\pm$ 0.10 \\
408089772 & 331.6978081 & \phantom{-}2.1919535 & 0.875 & 1.32 $\pm$ 0.10 & 10.91 $^{+0.05}_{-0.05}$ & 5.02 $\pm$ 0.37 & 0.75 $\pm$ 0.08 \\
408102456 & 330.9945490 & \phantom{-}2.2525661 & 0.880 & 1.18 $\pm$ 0.13 & 10.93 $^{+0.07}_{-0.09}$ & 5.34 $\pm$ 0.19 & 0.85 $\pm$ 0.12 \\
409065277 & 332.1249348 & \phantom{-}2.0521498 & 0.793 & 0.57 $\pm$ 0.05 & 10.91 $^{+0.09}_{-0.11}$ & 4.58 $\pm$ 0.36 & 0.80 $\pm$ 0.10 \\
409068570 & 332.5132463 & \phantom{-}2.0668140 & 0.754 & 1.10 $\pm$ 0.08 & 10.97 $^{+0.08}_{-0.10}$ & 5.69 $\pm$ 0.19 & 0.89 $\pm$ 0.08 \\
409085857 & 332.8420754 & \phantom{-}2.1449832 & 0.931 & 1.29 $\pm$ 0.11 & 11.10 $^{+0.09}_{-0.11}$ & 4.33 $\pm$ 0.10 & 0.91 $\pm$ 0.13 \\
410050835 & 333.6642052 & \phantom{-}1.9850561 & 0.952 & 1.12 $\pm$ 0.16 & 10.99 $^{+0.06}_{-0.07}$ & 5.01 $\pm$ 0.28 & 0.89 $\pm$ 0.12 \\
410067404 & 333.0045458 & \phantom{-}2.0621426 & 0.637 & 1.37 $\pm$ 0.07 & 10.90 $^{+0.08}_{-0.10}$ & 6.05 $\pm$ 0.27 & 1.10 $\pm$ 0.14 \\
410072423 & 332.8773225 & \phantom{-}2.0850862 & 0.829 & 0.87 $\pm$ 0.12 & 10.93 $^{+0.07}_{-0.09}$ & 5.77 $\pm$ 0.18 & 0.94 $\pm$ 0.09 \\
\hline
\end{longtable}
\endgroup
\end{document}